%% file: results.tex
\documentclass[10pt]{article}
\usepackage[onecolumn]{mnjen}

\usepackage{amstex}
\usepackage{epsfig}
\usepackage{amssymb}
\usepackage{eepic}

\newcommand{\be}{\begin{equation}}
	{\begin{equation}}%
{\end{equation}}

\input{defintns}

\title{Stability of Power-Law Disks II -- The Global Spiral Modes}
\author[N. W. Evans and J. C. A. Read]{N. W. Evans and J. C. A. Read 
\\ Theoretical Physics, Department of Physics, 1 Keble Rd, Oxford, OX1 3NP} 

\pubyear{1997}
 
\begin{document}
 
\maketitle
 
\begin{abstract}
\input{abstract}

\end{abstract}
 


\input{intro}

\input{bisymm}

\input{onearmed}

\input{triskele}

\input{tetraskele}

\input{selfcondisk}
\input{conc}

\bibliographystyle{mnbib}
\bibliography{biblio}

\setcounter{section}{0}
\setcounter{equation}{0}
\setcounter{figure}{0}
\setcounter{table}{0}
\appendix
\section{Axisymmetric Stability}\label{sec:axisymm}
\input{appa}

\setcounter{equation}{0}
\appendix
\section{The Symmetry of the Self-Consistent Disk}\label{sec:symmetry}
\input{appb}

\setcounter{equation}{0}
\appendix
\section{Neutral Modes of the Self-Consistent Disk}\label{sec:nmodes}
\input{appc}

\setcounter{equation}{0}
\setcounter{figure}{0}
\setcounter{table}{0}
\appendix
\section{Fastest Growing Modes of the Cut-Out Disks}\label{sec:fastest}
\input{appd}

\end{document}

%% file: defintns.tex
\def\modenergy{\left| \frac{\beta \Util^2 +\beta -2}{2\beta} \right|  } 
\def\betagammabig{  \frac{1}{\beta} + \frac{\gamma}{\beta} - \frac{\gamma}{2} }
\def\betagammasmall{  1/\beta + \gamma/\beta - \gamma/2 }

\def\auxint{ \mathcal{J} }

\def\fr#1#2{\textstyle {#1\over #2}\displaystyle}

\def\d2rdt2{{ {{d^2}R}\over {dt^2} }}
\def\half{{ {1 \over 2} }}

\def\Rtilc{ \tilde{R}_{\text{c}} }

\def\Util{ \tilde{U} }
\def\Ltil{ \tilde{L}_z }

\def\Qsing{ Q_{\text{s}} }

\def\Aimp{ A_{\text{imp}}}
\def\Ares{ A_{\text{res}}}

\def\alphamu{ \alpha_{\text{u}}}

\def\Sigeq{ \Sigma_{\text{eq}} }
\def\Sigres{ \Sigma_{\text{res}} }
\def\Sigmap{ \Sigma_{\text{p}} }

\def\fimp{ f_{\text{imp}} }
\def\psimp{ \psi_{\text{imp}} }
\def\vcirc{ v_{\text{circ}} }

\def\sigutil{ \tilde{\sigma}_u }

\def\sigutilmin{ \tilde{\sigma}_{u , \min}}

\def\kaptil{ \tilde{\kappa} }
\def\Omegap{ \Omega_{\text{p}} }
\def\Omtil{ \tilde{\Omega} }

\def\omtil{ \tilde{\omega} }
\def\stil{ \tilde{s} }

\def\Sm{ \mathcal{S}_m }

\def\Qlm{ Q_{lm}}
\def\Flm{ F_{lm}}
\def\lkapmom{ l\tilde{\kappa} + m\tilde{\Omega} }
\def\twopm{\frac{2+\beta}{2-\beta}}

\def\twomp{\frac{2-\beta}{2+\beta}}
\def\modterm{ \left| \frac%
	{  2 + 2\gamma - \beta \gamma }%
	{ \beta\Util^2 + \beta  - 2   } \right| }

\def\GammaFunctsombtpb{ \frac %
	{ \Gamma \left[ \half \left( 1-\beta \right)  \right]  %
         \Gamma \left[ \half \left( 2+\beta \right)  \right] } %
         { \Gamma \left[ \half \left( 1+\beta \right)  \right]  %
         \Gamma \left[ \half \left( 2-\beta \right) \right] } }

\def\Im {\text{Im} }
\def\Re {\text{Re} }

\def\spose#1{\hbox to 0pt{#1\hss}}
\def\lta{\mathrel{\spose{\lower 3pt\hbox{$\sim$}}
    \raise 2.0pt\hbox{$<$}}}
\def\gta{\mathrel{\spose{\lower 3pt\hbox{$\sim$}}
    \raise 2.0pt\hbox{$>$}}}

%% file: abstract.tex
\noindent 
This paper reports on the in-plane normal modes in the self-consistent
and the cut-out power-law disks.  Although the cut-out disks are
remarkably stable to bisymmetric perturbations, they are very
susceptible to one-armed modes. For this harmonic, there is no inner
Lindblad resonance, thus removing a powerful stabilising influence. A
physical mechanism for the generation of the one-armed instabilities
is put forward. Incoming trailing waves are reflected as leading waves
at the inner cut-out, thus completing the feedback for the
swing-amplifier. Growing three-armed and four-armed modes occur only
at very low temperatures.  However, neutral $m=3$ and $m=4$ modes are
possible at higher temperatures for some disks. The rotation curve
index $\beta$ has a marked effect on stability. For all azimuthal
wavenumbers, any unstable modes persist to higher temperatures and
grow more vigorously if the rotation curve is rising ($\beta <0$) than
if the rotation curve is falling ($\beta > 0$). If the central regions
or outer parts of the disk are carved out more abruptly, any
instabilities become more virulent.

The self-consistent power-law disks possess a number of unusual
stability properties.  There is no natural time-scale in the
self-consistent disk. If a mode is admitted at some pattern speed and
growth rate, then it must be present at all pattern speeds and growth
rates. Our analysis -- although falling short of a complete proof --
suggests that such a two-dimensional continuum of non-axisymmetric
modes does not occur and that the self-consistent power-law disks
admit no global non-axisymmetric normal modes whatsoever. Without
reflecting boundaries or cut-outs, there is no resonant cavity and no
possibility of unstable growing modes. The self-consistent power-law
disks certainly admit equiangular spirals as neutral modes, together
with a one-dimensional continuum of growing axisymmetric modes.

%% file: intro.tex
 \section{Introduction}

\noindent
The preceding paper in {\it Monthly Notices} has presented the
formidable machinery to permit direct solution for the in-plane normal
modes of the family of razor-thin power-law disks. The
results in this paper are based on a fast computer program that sets
this machinery working.

The entire spectrum of normal modes of a family of rigidly rotating
stellar dynamical disks was deduced by Kalnajs~\cite*{Kalnajs:1972} in
a bravura piece of mathematical analysis using \lq\lq only pencil,
paper and Legendre
polynomials\rq\rq~\cite{Toomre:1977}. Differentially rotating disks
are much harder and the results in the literature are sparse.
Kalnajs~\cite*{Kalnajs:1977} devised a general procedure for deducing
the normal modes of an axisymmetric stellar disk using matrix
mechanics in action-angle coordinates and he subsequently applied his
method to the isochrone disk~\cite{Kalnajs:1978}.  Over twenty years
have passed since Zang's~\cite*{Zang:1976} thesis on the scale-free
disk with a completely flat rotation curve, yet it remains the most
complete and instructive normal mode analysis available.  In fact --
perhaps daunted by the complexities of mode analyses -- many
researchers have looked for convenient short-cuts, such as gas
approximations~\cite{Bardeen:1975,Aoki:1979,BerLin:1996} or the use of
cold disks with softened gravity~\cite{Erickson:1974,Toomre:1981} or
the combination of N-body work with approximate analytic
treatments~\cite{Sell:1989}.  Our aim is to extend Zang's normal mode
work to all the scale-free disks with rising and falling rotation
curves, the power-law disks.  We are not alone in fixing attention on
these disks as suitably simple models for which global stability
analyses can be carried out. The gaseous power-law disks have
attracted the attentions of a number of
investigators~\cite{SE:1987,LKLB:1991,LBL:1993,SyTrem:1996}.  The
present paper gives a complete analysis of the global stability of the
cut-out and the self-consistent power-law disks of stars. In the
cut-out disks, stars close to the centre (and sometimes also at large
radii) are prevented from participating in the disturbance.  In the
self-consistent disks, all the stars are mobile.

Our study of the global stability of the cut-out power-law disks aims
to determine how much random motion is required to stabilise a disk to
modes of each angular harmonic.  A recurring theme of numerical
simulations throughout the seventies and eighties has been the
ferocity of the bar instability, even in disks that are safely stable
to axisymmetric
perturbations~\cite{Hohl:1971,AthanSell:1986,SellWilk:1993}.
Zang~\cite*{Zang:1976}, somewhat surprisingly, found that this is not
the case for the scale-free disk with a flat rotation curve. It is
bar-stable unless so cold that it is already susceptible to
axisymmetric modes. For each power-law disk, it is interesting to
locate the fastest growing modes and establish the physical origin of
the instabilities.  Are galaxies with rising rotation curves more or
less stable than galaxies with falling curves? What kinds of density
laws can be carved out to render a disk stable (or unstable)?  As
exact global stability analyses will remain scarce because of their
cumbersome nature, it is important to assess how rules-of-thumb --
such as Toomre's~\cite* {Toomre:1964} local criterion, or Ostriker and
Peebles'~\cite* {OstPeeb:1973} global criterion -- fare against exact
results.

The behaviour of the self-consistent disks must be very different from
that of the cut-out disk.  As originally pointed out by Kalnajs in
1974 (reported in Zang (1976)), all dependence on growth rate and
pattern speed can be factored out of the integral equation for the
normal modes. The same detail was noted independently
by~\longcite{LBL:1993} using an argument based on dimensional
analysis. Physical intuition suggests that as a disk is cooled below
the critical temperature, growing modes become possible.  But, the
symmetry of the scale-free disk forbids a preference for one growth
rate over another. At the critical temperature, the disk must admit
modes with all frequencies or none. Thus, there seem to be two
options: either there is a continuum of modes with all possible
pattern speeds and growth rates, or there are no normal modes
whatsoever. Do the scale-free, self-consistent power-law disks have
normal modes?  There are contradictory speculations on the answer to
this question~\cite{Zang:1976,LBL:1993}.

The paper first concentrates on the cut-out disks and examines their
global stability according to azimuthal wavenumber.  Bisymmetric
instabilities ($m=2$) are studied in Section 3. The one-armed modes
($m=1$) are examined in Section 4, while the other azimuthal harmonics
($m=3,4$) in Section 5. Finally, Section 6 considers the
self-consistent disk in the light of the insights gained in the study
of cut-out disks. Before all this, however, we begin with a short
introductory Section 2, which introduces some common notation and
themes for the rest of the paper.

%% file: bisymm.tex
\section{Theoretical Preamble}
Before discussing stability to any non-axisymmetric modes, it is
helpful to start with a brief discussion of axisymmetric
stability. This provides part of the intellectual framework needed to
understand the non-axisymmetric results.  For short wavelength,
axisymmetric modes, ~\longcite{Toomre:1964} showed that local
stability requires that the velocity dispersion exceed a critical
value $\sigutilmin$. First, let us make the convenient abbreviation
\be
     L(\beta) = 	\GammaFunctsombtpb,
\end{equation}
For the power-law disks, the critical velocity dispersion for
stability to axisymmetric modes is
\be
	\sigutil > \sigutilmin = \frac{3.3582830} {2 \pi L(\beta)
				\sqrt{2-\beta}}
				\frac{\Sigma_{\text{active}}}{\Sigeq}
				\label{eq:localsigmin},
\end{equation}
where $\Sigma_{\text{active}}$ is the active surface density or the
density spared by the cut-out function and $\Sigeq$ is the surface
density implied by Poisson's equation.  For the cut-out disks,
$\sigutilmin$ is a function of radius, although for the
self-consistent disk it is not.  We shall often relate the velocity
dispersion $\sigutil$ in both self-consistent and cut-out disks to the
minimum velocity dispersion necessary to ensure local axisymmetric
stability in the self-consistent disk. This ratio is referred to as
$\Qsing$
\be
	\Qsing 
	= \sigutil 
	\frac{2 \pi  \sqrt{2-\beta} }{3.3582830} 
	L(\beta)
	\label{eq:Qsing}.
\end{equation}
Local theory tells us that $\Qsing \ge 1$ is sufficient to stabilise
against short wavelength axisymmetric modes. This result also holds
good for stability against global axisymmetric modes, as has been
found before by others~\cite{Zang:1976,LKLB:1991} and is briefly
demonstrated anew in Appendix~\ref{sec:axisymm}. 
\begin{table}
\begin{center}
\begin{tabular}
{|l|l|l|l|}
\hline
$\beta$
&$L(\beta)$
&$m X$
&$\sigutilmin$
\\ \hline
$-0.75$
& $0.233406$
& $0.641867$
& $1.380888$
\\ 
$-0.5$ 		
& $0.456947$ 	
& $1.142366$		
& $0.739779$	
\\ 
$-0.25$ 		
& $0.700219$
& $1.575492$ 
& $0.508877$
\\
$\null$ 		
& $\null$
& $\null$ 
& $\null$
\\ 
$0.0$	
&$1.000000$
&$2.000000$
&$0.377940$
\\
$\null$ 		
& $\null$
& $\null$ 
& $\null$
\\ 
$0.25$
& $1.428125$ 
& $2.49929$
& $0.282912$
\\ 
$0.5$
& $2.188440$
& $3.282660$
& $0.199415$
\\ 
$0.75$
& $4.284375$
& $5.355469$ 
& $0.111582$ 
\\ \hline
\end{tabular}

\caption{This table shows the variation of some useful stability 
indicators with the rotation curve index $\beta$ of the self-consistent
power-law disks. Here, $\sigutilmin$ is the minimum velocity
dispersion (measured relative to the local circular speed) needed to
suppress axisymmetric Jeans instabilities entirely. $X$ is the ratio
of the circumferential wavelength of an $m$-armed disturbance to
the longest Jeans unstable wavelength $\lambda_{\rm crit}$ in a cold
disk.}
\end{center}

\end{table}
\medskip

\noindent
Toomre (1964) showed that rotation alone stabilises axisymmetric modes
with wavelengths in excess of $\lambda_{\rm crit}$
\be
\lambda_{\rm crit} = {4 \pi^2 G \Sigma \over \kappa^2} =
{2 \pi R \over (2-\beta) L(\beta)}
{\Sigma_{\text{active}}\over \Sigeq}.
				\label{eq:lambdacrit}
\end{equation}
Equivalently, lengthscales larger than $\lambda_{\rm crit}$ are stable
to Jeans instabilities even in a cold disk. It is often helpful to
discuss non-axisymmetric instabilities in terms of a dimensionless
ratio $X$ (e.g., Toomre 1981).  This is the ratio of the
circumferential wavelength of an $m$-armed disturbance $\lambda$ to
$\lambda_{\rm crit}$, i.e.,
\be
X = \frac{\lambda}{\lambda_{\rm crit}} = {2 \pi R \over m \lambda_{\rm
crit}} = \frac{(2-\beta)L(\beta)}{m} 
{\Sigeq\over\Sigma_{\text{active}}}.
				\label{eq:xdef}
\end{equation}
Table 1 shows how these helpful stability parameters vary for the
self-consistent power-law disks ($\Sigeq = \Sigma_{\text{active}}$).
The table makes the point that the properties of the disks change
quickly as one moves away from the reference Toomre-Zang disk ($\beta
=0$). For example, $\lambda_{\rm crit}$ increases by a factor of
$1.75$ on moving from $\beta =0$ to $\beta = -0.5$ and decreases by a
factor of $1.64$ on passing from $\beta =0$ to $\beta = 0.5$.  The
densities of any clumps formed simply through axisymmetric Jeans
instabilities are almost a factor of three larger on moving from
$\beta = 0.5$ to $\beta = -0.5$.  Relative to the local circular
speed, $\sigutilmin$ roughly doubles or halves on making the same
journeys.  This table already suggests that the disks with $\beta <0$
will be harder to stabilise against Jeans instabilities and their kin
than the disks with $\beta >0$.  The swing-amplifier is at its
fiercest when $1\lta X \lta 2$ and becomes weak once $X>3$ (see Toomre
1981). So, Table 1 cautions us that the one-armed ($m=1$) modes
especially in the disks with rising rotation curves ($\beta <0$) will
be particularly troublesome to stabilise.
%
%
\section{Bisymmetric Modes}\label{sec:m2chapter}
\noindent
This section provides the results of the stability analysis to global
bisymmetric perturbations. From an observational standpoint, these are
the most important, as some $30 \%$ of disk galaxies are strongly
barred with perhaps a further $30 \%$ showing evidence of weaker
barring~\cite{SellWilk:1993}. There are even galaxies known which are
unbarred in the optical but are barred in the infrared.  Approximate
bisymmetry is a characteristic of most disk galaxies.
%
%
\subsection{The Behaviour of the Eigenvalues}
\subsubsection{The Effect of Growth Rate and Pattern Speed}
Let us recall that the mathematical eigenvalue $\lambda$ is the ratio
of the response to the imposed density transforms (see Section 4.3 of
the preceding paper). Mathematically, it is defined by $\Ares (\alpha)
= \lambda \Aimp (\alpha )$. If the modulus of the complex mathematical
eigenvalue $|\lambda|$ is greater than unity, the response is more
vigorous than the imposed perturbation. If $|\lambda|<1$, the response
is less than the original disturbance. When the mathematical
eigenvalue is unity, the response is identical in magnitude and phase
to the imposed disturbance. This is a self-consistent mode.  To
discover global modes, it is important to understand the position of
the eigenvalue in the complex plane and to learn what choices of
growth rate, pattern speed and temperature can bring it to the point
$(1,0)$.  Cool disks are (usually) less stable than warm ones. By
increasing the temperature sufficiently, the modulus of the
mathematical eigenvalue shrinks so that no choice of growth rate $s$
and pattern speed $\Omegap$ enables it to attain the value $(1,0)$.
On physical grounds, slowly-growing disturbances are expected to be
more easily excited than faster ones. Thus, the modulus of the
mathematical eigenvalue is greater for smaller growth rates. The
pattern speed has a twofold effect, as it controls the phase of the
response and determines the position of the co-rotation radius
$R_{\text{CR}}$ and the inner and outer Lindblad resonances,
$R_{\text{ILR}}$ and $R_{\text{OLR}}$.  For the power-law disks, these
crucial radii occur at:
\begin{xalignat}{3}
	\frac{R_{\text{ILR}}}{R_0} &= \left( \frac{m - \sqrt{2-\beta}}
	{m\Omtil_{\text{p}}} \right) ^{\frac{2}{2+\beta}},
&\quad
	\frac{R_{\text{OLR}}}{R_0} &= \left( \frac{m + \sqrt{2-\beta}}
	{m\Omtil_{\text{p}}} \right) ^{\frac{2}{2+\beta}},
&\quad
	\frac{R_{\text{CR}}}{R_0} &= \left(
\frac{1}{\Omtil_{\text{p}}}
          \right)^{\frac{2}{2+\beta}}.
	\label{eq:lindblad}
\end{xalignat}
These formulae are given for general azimuthal wavenumber $m$.
Henceforth, the emphasis in this section is on $m=2$.  Any modes are
generally confined to the region between the two Lindblad resonances.
For $\Omtil_{\text{p}} \gtrsim 1+\fr12 \sqrt{2-\beta}$, the outer
Lindblad resonance lies inside the inner cut-out at $R=R_0$. The
disturbance is restricted to a part of the disk where there is not
much matter free to respond. The response, and hence the mathematical
eigenvalue, is expected to be small for large pattern speeds.  For
$\Omtil_{\text{p}} \lesssim 1-\fr12 \sqrt{2-\beta}$, the inner
Lindblad resonance lies beyond the inner cut-out. Now, the presence of
the inner cut-out has a minimal effect on the response, since the
density experienced is similar to that in the self-consistent
disk. Each imposed logarithmic spiral then excites only a pure
logarithmic spiral in response and so the eigenvalue is expected to be
real.  In a doubly cut-out disk, with sufficiently low pattern speed,
the inner Lindblad resonance can fall beyond the outer cut-off at
$R=R_{\text{c}}$.  In this case, the mathematical eigenvalue is again
expected to be very small.

\begin{figure}
	\begin{center}
		\epsfig{file=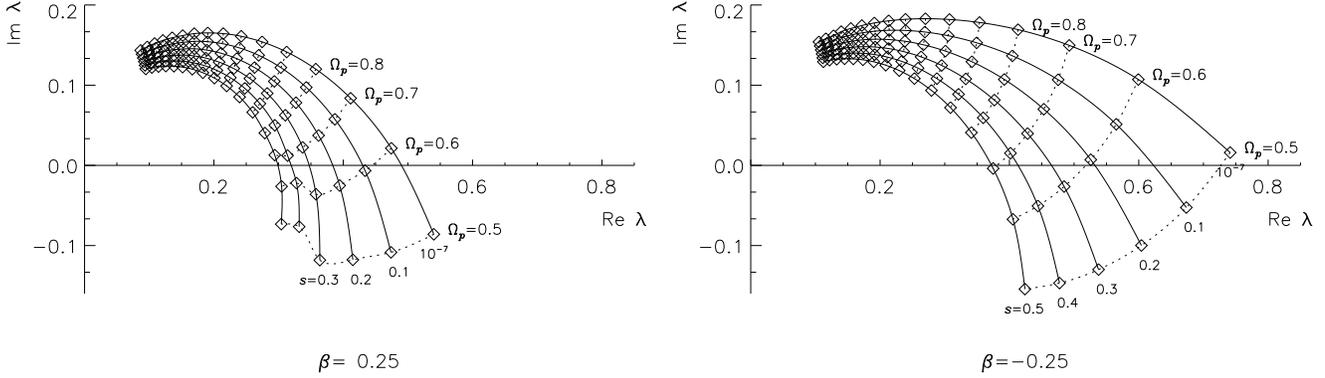,width=\textwidth,height=0.3\textwidth}
		\caption[The dependence of the largest eigenvalue on
		growth rate and pattern speed, for an inner-cut-out
		disk with $\beta=\pm0.25$]{The dependence of the
		largest mathematical eigenvalue on growth rate and
		pattern speed for bisymmetric disturbances in a inner
		cut-out disk with $N=2$, $\Qsing=1$ and
		$\beta=\pm0.25$. [The solid lines show curves of
		constant $s$ at intervals of 0.1 from $s=0^+$ to
		$s=0.5$. The dotted lines show curves of constant
		$\Omegap$ at intervals of 0.1 from $\Omegap=0.5$ to
		$\Omegap=2.0$. For $\beta=0.25$, $\sigutil = 0.283$
		and $\gamma=11.0$; for $\beta=-0.25$, $\sigutil =
		0.509$ and $\gamma=3.36$].
\label{fig:OmegapLoop113_b0pm25_Q1}}
		\end{center}
\end{figure}

Fig.~\ref{fig:OmegapLoop113_b0pm25_Q1} shows the effect of growth rate
and pattern speed on the mathematical eigenvalue for $N=2$ disks with
$\beta = \pm 0.25$. The temperature of each disk is chosen so that it
is locally stable to axisymmetric perturbations, i.e., $\Qsing=1$. 
The curves have the general behaviour expected from our theoretical
discussion. As the pattern speed is increased beyond $\sim 1$, the
modulus of the mathematical eigenvalue declines. For large $\Omegap$,
the eigenvalue turns towards the origin and the growth rate becomes
smaller. The disk responds to the perturbation rather weakly, since
most of the mass has been removed from the central region.  Now, the
mathematical eigenvalue depends on growth rate $s$ and pattern speed
$\Omegap$ through the single complex frequency $\omega=m\Omegap+is$.
Since the kernel is analytic, the mathematical eigenvalue $\lambda$ is
an analytic function of $\omega$.  As realised by
Zang~\cite*{Zang:1976}, this means that $\lambda(\omega)$ gives a
conformal mapping of the upper half of the complex $\omega$-plane
(since the growth rate is positive) to the complex
$\lambda$-plane. Conformal mappings preserve the angles of
infinitesimal polygons.  So, curves of constant $s$ and constant
$\Omegap$ are orthogonal.  Fig.~\ref{fig:OmegapLoop113_b0pm25_Q1}
shows a problem which can occur. For $\beta=0.25$, there is a
discontinuity near ($\Omegap=0.5$, $s=0.4$), even though $\lambda$ is
expected to change smoothly. The discontinuity arises because here the
second largest eigenvalue overtakes the previously largest one, so
that the same eigenvalue is not being followed.

\begin{figure}
	\begin{center}
		\epsfig{file=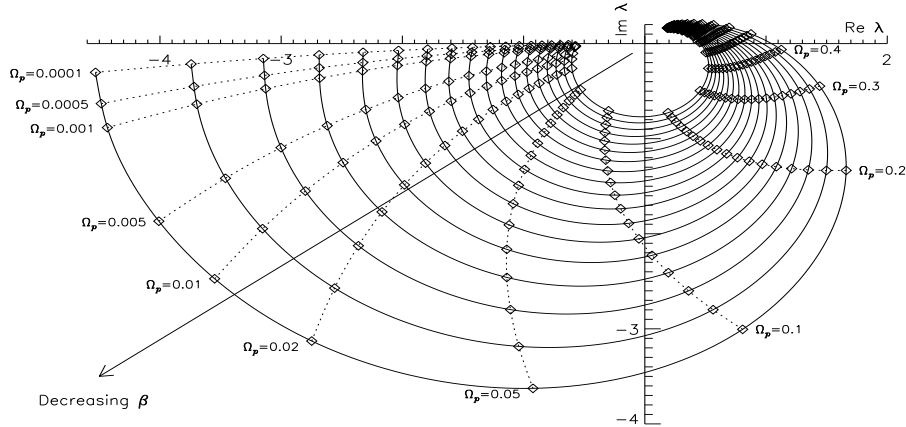,width=0.7\textwidth,height=0.3\textwidth}
		\caption[Marginal eigenvalue curves for inner cut-out
		disks with different values of $\beta$] {Marginal
		eigenvalue curves for bisymmetric disturbances in
		inner cut-out disks with $N=2$ and values of the
		rotation curve index from $\beta=-0.5$ to $\beta=0.35$
		in steps of 0.05.  The velocity dispersion for each
		$\beta$ is that corresponding to local axisymmetric
		stability ($\Qsing=1$).  [The plot shows the largest
		mathematical eigenvalues for vanishing growth rate,
		$s=10^{-7}$, and 20 values of the pattern speed:
		$\Omegap = 0.0001$, 0.0005, 0.001, 0.005, 0.01, 0.02,
		0.05, 0.1, 0.2, 0.3, $\dots$ 1.2, 1.3.]
		\label{fig:MarginalManyBeta_Q1_m2}} \end{center}
\end{figure}

What are the values of the growth rate, pattern speed and temperature
for which the eigenvalue is unity?
Fig.~\ref{fig:OmegapLoop113_b0pm25_Q1} already suggests that, for
$\Qsing=1$ disks with $\beta$ in the range $-0.25$ to $0.25$, no
choice of growth rate and pattern speed results in an eigenvalue equal
to unity. For each $\beta$, the curves of constant growth rate cross
the real axis well to the left of unity. From our theoretical
discussion, the mathematical eigenvalue is likely to continue towards
the origin as the pattern speed increases still further. Assuming
that, at any pattern speed, the marginal eigenvalue curve has the
largest modulus and that the curves of constant $s$ do not cross, it
seems improbable that any solutions can exist at pattern speeds lower
than those shown.  Fig.~\ref{fig:OmegapLoop113_b0pm25_Q1} plots the
behaviour of the dominant eigenvalue, i.e., that which has the largest
modulus at a particular pattern speed and growth rate. For the range
of pattern speeds investigated so far, the modulus of the dominant
eigenvalue has not reached unity, and so none of the lower eigenvalues
can yield a mode.  Let us also observe from
Fig.~\ref{fig:OmegapLoop113_b0pm25_Q1} that as $\beta$ is reduced from
$0.25$ to $-0.25$, the set of curves tilts upwards, crossing the real
axis at higher values, and thus coming closer to possessing a unit
eigenvalue. The disks with rising rotation curves (negative $\beta$)
are less securely stable than their relatives with falling rotation
curves (positive $\beta$).  Fig.~\ref{fig:MarginalManyBeta_Q1_m2}
shows marginal eigenvalue curves for several different values of
$\beta$ in the range $-0.5$ to $+0.35$.  The curves are drawn for
$\Qsing=1$, so all the disks shown are already stable to axisymmetric
perturbations. The disks with rising rotation curves have higher
velocity dispersions than those with falling rotation curves, but even
so they are more prone to bisymmetric disturbances.  As $\beta$ is
decreased, the magnitude of the marginal eigenvalue at a given pattern
speed increases. Eigenvalues in disks with rising rotation curves
cross the real axis at higher values than for their cousins with
falling rotation curves.
For values of $\beta < -0.470$, $\Qsing=1$ is insufficient to
stabilise the disk to $m=2$ perturbations. Disks whose rotation curves
rise faster than this succumb to bar-like modes, even if stable to
axisymmetric modes.
%
%
\subsubsection{The Effect of the Temperature and the Inner 
Cut-Out\label{sec:m2InnerCutOut}}
\begin{figure}
	\begin{center}
		\epsfig{file=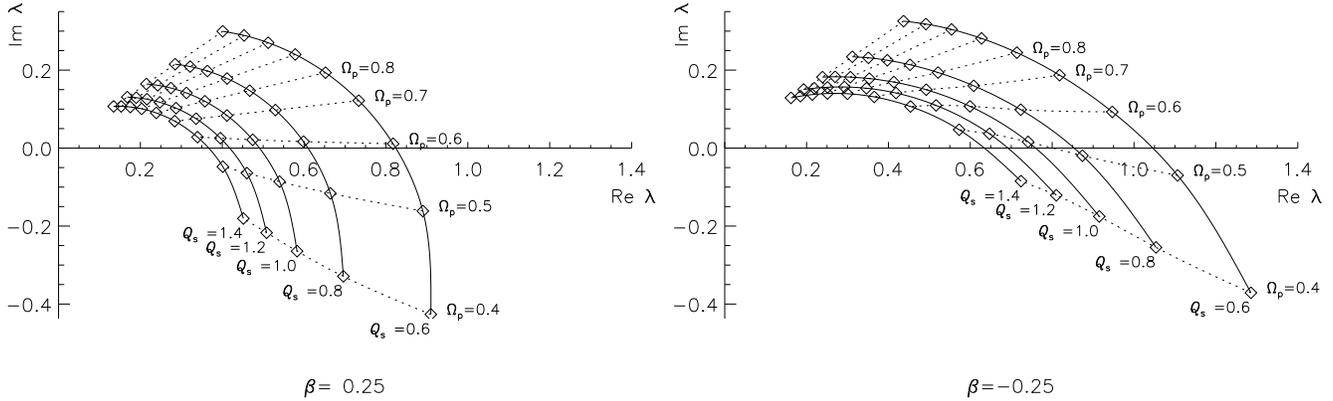,width=\textwidth,height=0.3\textwidth}
		\caption [Dependence of the largest mathematical
		eigenvalue on temperature and pattern speed, for
		vanishing growth rate and $\beta=0.25$] {The
		dependence of the largest mathematical eigenvalue on
		temperature and pattern speed for inner cut-out disks
		with $\beta=\pm 0.25$. Curves are plotted for 
		bisymmetric perturbations with vanishing growth rate in
		a disk with cut-out index $N=2$. The temperature used
		for each curve is expressed in terms of $\Qsing$. [For
		$\beta=0.25$, $\sigutil = 0.283 \Qsing$; for
		$\beta=-0.25$, $\sigutil = 0.509
		\Qsing$].\label{fig:Fixeds_2_s000_b0pm25}} \end{center}
\end{figure}

Raising the velocity dispersion $\sigutil$ moves the marginal
stability curves closer to the origin, while lowering it moves them
further out. At some value $\sigutilmin$, the marginal stability curve
will pass through (1,0); this is the temperature at which the disk is
just stable to bisymmetric
perturbations. Fig.~\ref{fig:Fixeds_2_s000_b0pm25} shows how the
temperature affects the marginal stability curve.  The eigenvalue
depends separately on the pattern speed and temperature, so the
mapping from the ($\Omegap$-$\Qsing$) plane to the complex $\lambda$
plane is not conformal. This is evident from
Fig.~\ref{fig:Fixeds_2_s000_b0pm25}, where lines of constant $\Omegap$
and constant $\Qsing$ do not intersect at right-angles.  Again, this
demonstrates that power-law disks with falling rotation curves are
securely stable to bisymmetric perturbations. Even when the velocity
dispersion is only 60\% of that needed for axisymmetric stability, the
$\beta=0.25$ disk admits no $m=2$ modes.

\begin{figure}
	\begin{center}
		\epsfig{file=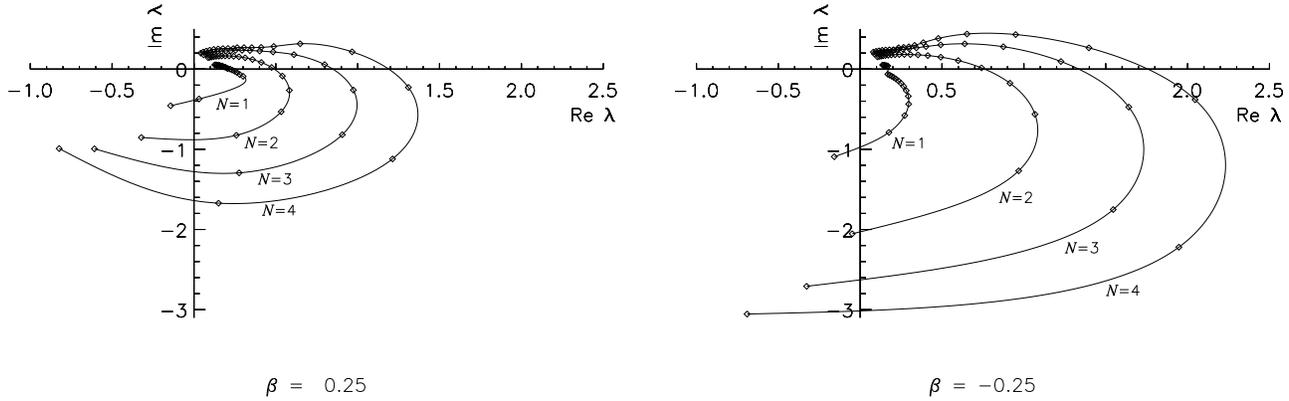,width=\textwidth,height=0.3\textwidth}
		\caption[Marginal eigenvalue curves for disks with
		different inner cut-out indices] {The dependence of
		the largest mathematical eigenvalue on pattern speed,
		for different inner cut-out indices $N$. Curves are
		plotted for bisymmetric perturbations with vanishing
		growth rate in an inner cut-out disk with $\beta=\pm
		0.25$. The curves are labelled with the value of the
		inner cut-out index $N$. [In each case $\Qsing=1$. The
		diamonds show the results for $\Omegap=0.1$, 0.2, 0.3
		... 2.0.]
\label{fig:EffectOfNi}
}
	\end{center}
\end{figure}

The inner cut-out strongly influences the stability of the disk (c.f.,
Zang 1976; Toomre 1977). To illustrate this, Fig.~\ref{fig:EffectOfNi}
shows the marginal stability curves in disks with different $\beta$
and different values of the inner cut-out index $N$.
Increasing the cut-out index $N$ makes the disk much more susceptible
to instabilities. Curves with higher $N$ cross the real axis at
successively higher values of $\lambda$. Disks with $N=1$ are highly
stable to $m=2$ perturbations.  To understand why the inner cut-out
has such a profound effect on the stability of the disks, let us
consider the position of the Lindblad resonances. For low values of
the pattern speed, the inner Lindblad radius lies outside the inner
cut-out radius $R_0$. Incoming waves are absorbed by the inner
Lindblad resonance, which thus damps disturbances in the disk and
reduces the possibility of a mode. However, once $\Omegap$ exceeds
$\sim 0.25$, the inner Lindblad resonance moves inside the inner
cut-out radius.  If the cut-out is gentle ($N=1$), the incoming waves
pass through it, reach the inner Lindblad resonance and are
absorbed. If however the cut-out is sufficiently sharp (as for larger
cut-out indices), it presents a barrier which reflects the incoming
trailing waves. Instead of reaching the inner Lindblad resonance and
being absorbed, they are reflected back as outgoing leading waves.
The exceptional stability of $N=1$ disks is mostly due to the fact
that the surface density is disturbed much less abruptly than for
higher cut-out indices, presenting less of a reflective barrier to
incoming waves. Additionally, there is a much greater surface density
within the inner cut-out when $N=1$; the active surface density
diverges at the origin when the rotation curve is falling, and even
when the rotation curve is rising, the surface density remains large
to much smaller radii. By contrast, for $N>1$, the active surface
density rapidly falls to zero within $R_0$. Waves are thus suppressed
by simply having no medium through which to propagate.
%
%
\subsection{The Marginal Modes\label{sec:Marginal}}
\begin{figure}
	\begin{center}
		\epsfig{file=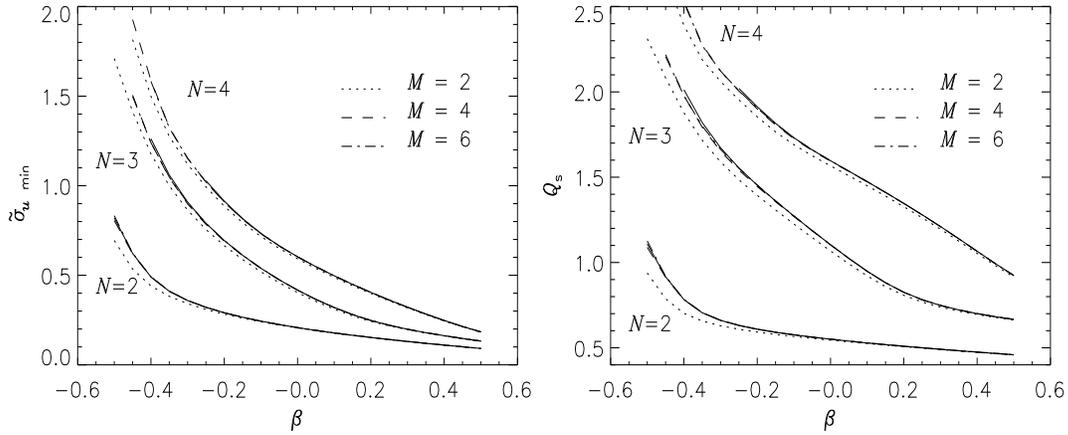,width=0.8\textwidth,height=0.35\textwidth}
		\caption[Minimum temperature for global stability
		plotted against $\beta$, for doubly cut-out disks]
		{Minimum temperature for bisymmetric stability plotted
		against rotation curve index $\beta$ for disks with
		various cut-out functions. The left-hand plot shows
		the minimum velocity dispersion $\sigutil$; the
		right-hand one the same data presented in terms of the
		stability parameter $\Qsing$. The solid lines indicate
		the results for inner cut-out disks with $N=1$, 2, 3,
		4. The broken lines indicate the corresponding doubly
		cut-out disks with $M=2$, 4, 6 and $\Rtilc=10$.
		\label{fig:m2margsiggam}} \end{center}
\end{figure}

\begin{figure}
	\begin{center}
		\epsfig{file=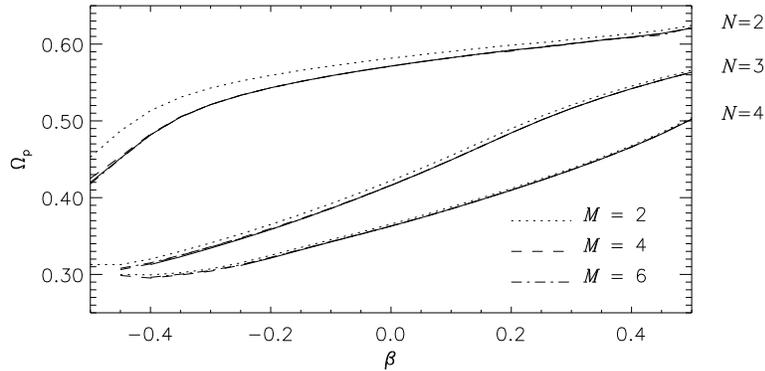,width=0.6\textwidth,height=0.3\textwidth}
		\caption[Critical pattern speed plotted against
		$\beta$, for doubly cut-out disks] {Critical pattern
		speed $\Omegap$ plotted against rotation curve index
		$\beta$ for $m=2$ modes in disks with various cut-out
		functions. The solid lines are labelled with the inner
		cut-out index $N$. For each $N$, different outer
		cut-out indices $M$ are plotted.
\label{fig:m2margrot}} \end{center}
\end{figure}

The marginal modes are those with vanishingly small growth rate. How
cold does the disk have to be for this transition from stability to
instability to occur? This question can be answered by finding the
temperature for which the marginal eigenvalue curve passes through
$(1,0)$.  Fig.~\ref{fig:m2margsiggam} shows the minimum temperature
needed for bisymmetric stability as a function of $\beta$. The solid
lines show the results for the inner cut-out disks with $N=2$, 3,
and 4. The broken lines show results for the corresponding doubly cut-out
disks with $M=2$, 4, 6 and $\Rtilc=10$.
Curves with different $N$ lie well apart. The more sharply the core is
cut out, the more unstable the disk becomes.  For each $N$, the curves
with different $M$ lie very close together. Even with this low
truncation radius, the outer cut-out function makes little difference
to the temperature necessary for stability. With a larger $\Rtilc$,
the doubly cut-out curves are scarcely distinguishable from those with
only an inner cut-out.  The right-hand plot in
Fig.~\ref{fig:m2margsiggam} can be used to compare the relative
stability of the various cut-out disks to $m=0$ and $m=2$
perturbations.  The $N=2$ disk is less prone to bisymmetric than to
axisymmetric instabilities over almost all $\beta$. For $N=3$, the
relative tendency to $m=2$ and $m=0$ disturbances depends on
$\beta$. Roughly speaking, $N=3$ disks with rising rotation curves are
more susceptible to $m=2$ disturbances whereas disks with falling
rotation curves are more susceptible to $m=0$. Conversely, disks with
$N=4$ are much more prone to bisymmetric disturbances over almost all
$\beta$. Only when the rotation curve is steeply falling does the
$m=0$ mode become harder to stabilise.  It is perhaps surprising that
it is the lowest outer cut-out index ($M=2$) which shows the largest
departure from the inner cut-out curves. In this instance, it seems
that tapering the disk gently at large radius has more of an effect
than truncating it abruptly. This is presumably because for lower $M$,
the effect of the outer cut-out is felt further in towards the centre
of the disk. For higher $M$, the outer cut-out is sharp, but occurs
entirely outside the outer Lindblad resonance. Disks with a gentle
outer cut-out are more stable than those where the cut-out is sharper.
The critical pattern speed obtained for these marginal modes is shown
in Fig.~\ref{fig:m2margrot}. For all cut-out functions and $\beta$,
the pattern speed is confined to a fairly narrow range, roughly 0.3 to
0.6. Over a wide range in $\beta$, the pattern speed varies roughly
linearly with $\beta$.

\begin{figure}
	\begin{center}
		\epsfig{file=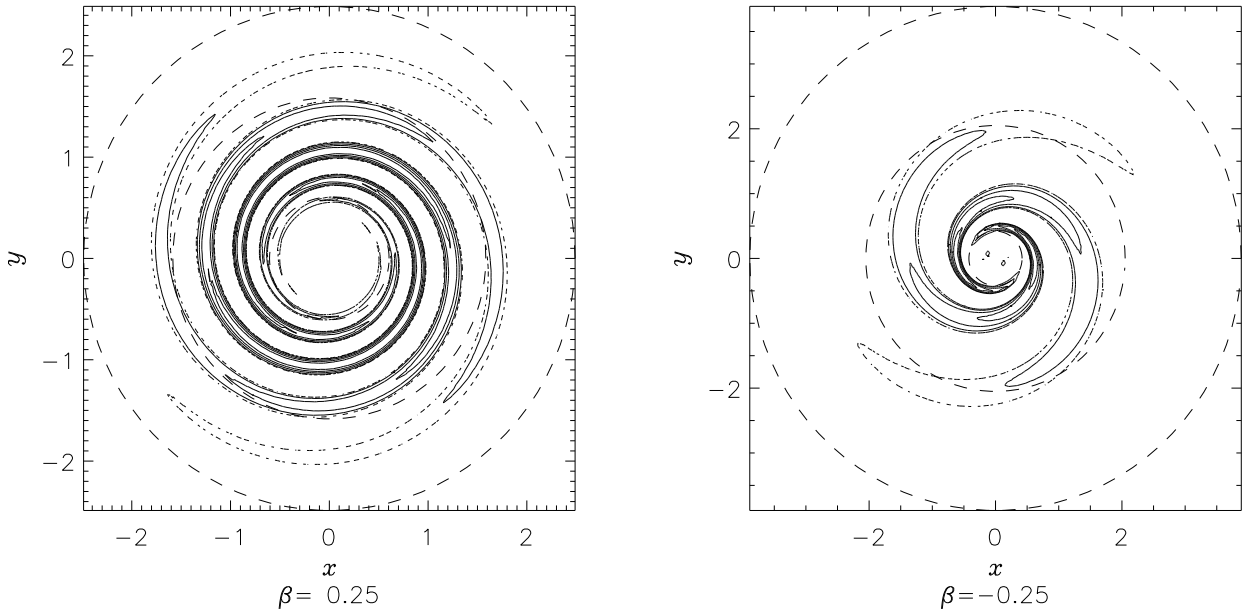,width=\textwidth,height=0.5\textwidth}
		\caption [Density of marginally stable modes in inner
		cut-out disks with $N=2$, $\beta=\pm 0.25$.]  {Density
		contour plot of a marginally stable mode in inner
		cut-out disks with $N=2$.  The solid lines mark 10\%,
		20\%, 40\%, 60\% and 80\% of the maximum density. The
		dotted lines show the nodes, where the perturbation
		density vanishes.  The Lindblad resonances and the
		co-rotation radius are indicated by dashed
		circles. [The left-hand plot is for $\beta=0.25$, for
		which the marginally stable mode has $\Qsing=0.500$,
		$\gamma=48.4$, $\sigutil=0.142$, $\Omegap=0.596$ and
		$\alphamu\approx 20.4$.  The right-hand plot is for
		$\beta=-0.25$, for which the marginally stable mode
		has $\Qsing=0.631$, $\gamma=9.19$, $\sigutil=0.321$,
		$\Omegap=0.533$ and $\alphamu\approx 8.4$].
		\label{fig:Mode_N2_b0pm25_Qmarg}}
\end{center}
\end{figure}

\begin{figure}
	\begin{center}
		\epsfig{file=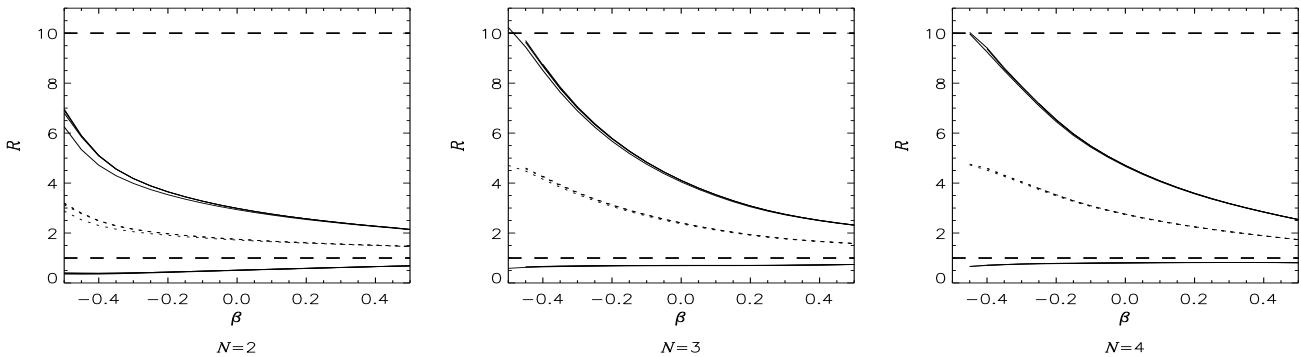,width=\textwidth,height=0.3\textwidth}
		\caption[Position of resonances for marginal
		modes]{Position of resonances for marginal $m=2$ modes
		in cut-out disks with different rotation curve index
		$\beta$. The dashed horizontal lines show the position
		of the inner and outer cut-out radii. The pattern
		speed corresponding to the marginal mode in each disk
		(Fig.~\ref{fig:m2margrot}) is used to calculate the
		the resonance radii. The solid lines mark the Lindblad
		resonances; the dotted line marks co-rotation. [The
		three plots show results for $N=2$, 3 and 4; each plot
		shows data for doubly cut-out disks with $M=2$, 4, 6,
		$\Rtilc=10$].
                \label{fig:m2_marginal_RILR}}
		\end{center}
\end{figure}

Let us look at the shapes of some of these marginally stable modes.
The contour plots in Fig.~\ref{fig:Mode_N2_b0pm25_Qmarg} show the form
of the modes for $\beta = \pm 0.25$ in inner cut-out disks with $N=2$.
For falling rotation curves, the modes are more tightly wound than for
rising rotation curves. They extend beyond co-rotation up to the outer
Lindblad resonance, whereas in disks with rising rotation curves the
modes are concentrated between the inner Lindblad resonance and the
co-rotation radius. The modes are also more tightly wound for disks
with gentle cut-outs. These effects are caused by the higher
temperatures needed for marginal modes when the rotation curve is
rising. If a growing mode exists, then the stars do not have enough
random motion to move entirely out of a density enhancement before it
has grown significantly. As the temperature increases, the stars have
more random motion, and the spiral modes can therefore be more loosely
wound. As shown in Fig.~\ref{fig:m2margrot}, the pattern speed of
modes increases with $\beta$ and is greater for disks with lower
cut-out indices. Since the Lindblad resonances move inwards as the
pattern speed increases, the pattern is concentrated within a smaller
radius for $\beta=0.25$ than for $\beta=-0.25$, and similarly for the
lower cut-out indices. Fig.~\ref{fig:m2_marginal_RILR} shows the
position of the Lindblad and co-rotation resonances for modes in
cut-out disks.
It is significant that, for every value of $\beta$, the pattern speed
is large enough to push the inner Lindblad resonance just inside the
inner cut-out. The disk does not admit slowly-rotating growing modes
for which the resonance would emerge beyond the cut-out. The resonance
is allowed nearer the cut-out when the cut-out is sharper, and hence a
more efficient reflector.  Similarly, on examining modes in an
unstable disk (Read 1997), the fastest-growing modes have the
largest pattern speeds. As the pattern speed decreases, allowing the
inner Lindblad resonance to move out closer to the cut-out radius, the
growth rate slows.

\subsection{The Ostriker-Peebles Criterion}
Ostriker and Peebles~\cite*{OstPeeb:1973} suggested that disk galaxies
are stable to bar-like modes only when the ratio of total rotational
energy $T$ to total gravitational energy $|W|$ is less than 0.14. The
energies $W$ and $T$ for the self-consistent disk are given in Section
2 of Paper I. We thus deduce the ratios as~\cite{Evans:1994}:
\begin{xalignat}{2}
	\frac{T}{|W|} & = \frac
	{\Gamma^2\left[ 1+\frac{\gamma}{2} \right]
	 \Gamma^2\left[ 2+\frac{1}{\beta}+\frac{\gamma}{\beta} \right]}
	{\beta \Gamma^2\left[ \frac{1}{2}+\frac{\gamma}{2} \right]
	 \Gamma^2\left[ \frac{5}{2}+\frac{1}{\beta}+\frac{\gamma}{\beta}\right]} ,
& \quad
	\beta >0,
	\label{eq:TWposbet}
\\
	\frac{T}{|W|} & = \frac
	{\Gamma^2\left[ 1+\frac{\gamma}{2} \right]
	 \Gamma^2\left[ -\frac{3}{2}-\frac{1}{\beta}-\frac{\gamma}{\beta} \right]}
	{-\beta \Gamma^2\left[ \frac{1}{2}+\frac{\gamma}{2} \right]
	 \Gamma^2\left[ -1-\frac{1}{\beta}-\frac{\gamma}{\beta}\right]} ,
& \quad
	\beta <0,
	\label{eq:TWnegbet}
\\
	\frac{T}{|W|} & = \frac{1}{1+\gamma}
	\frac
	{\Gamma^2\left[ 1+\frac{\gamma}{2} \right] }
	{ \Gamma^2\left[ \frac{1}{2}+\frac{\gamma}{2} \right]} ,
& \quad
	\beta = 0.
\end{xalignat}
Note that the $\beta=0$ result can be derived from the values for
$\beta \ne 0$ using a property of the gamma function
~\cite[8.328.2]{Grad:Rhyz}.

\begin{figure}
	\begin{center}
		\epsfig{file=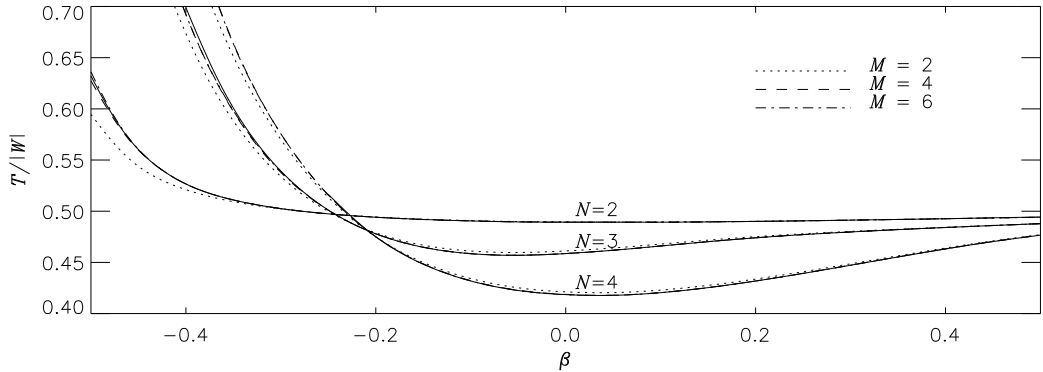,width=0.8\textwidth,height=0.3\textwidth}
		\caption[Comparison with the Ostriker-Peebles
		criterion] {Comparison of the results for the cut-out
		disks with the Ostriker-Peebles criterion. The plot
		shows the ratio of total rotational energy to total
		potential energy necessary for marginal stability to
		bar modes. The solid lines are labelled with the
		appropriate inner cut-out index $N$. For each $N$,
		different outer cut-out indices $M$ are plotted, as
		shown in the key.
\label{fig:OstrikerPeebles}
} 
	\end{center}
\end{figure}

Fig.~\ref{fig:OstrikerPeebles} shows the ratio $T/|W|$ calculated for
the self-consistent disk (eqs.~\eqref{eq:TWposbet}
and~\eqref{eq:TWnegbet}). The anisotropy parameter used at each
$\beta$ is that needed for marginal stability to bisymmetric modes in
the cut-out disk.
Note that $T/|W|$ can exceed 0.5, since the standard form of the
virial theorem does not apply to the power-law disks, which are of
infinite extent.  For all the disks, the ratio $T/|W|$ is much greater
than the value of 0.14 suggested by Ostriker and Peebles.  This
demonstrates that the Ostriker-Peebles criterion is not useful in
predicting the stability of the power-law disks.  However, the
temperatures used in Fig.~\ref{fig:OstrikerPeebles} are derived for
the cut-out disks.  As the self-consistent and cut-out disks have
different stability properties, perhaps the value of $T/ |W|$ should
be calculated for the cut-out disks.  In this case, the density
$\Sigma$ and streaming velocity $\langle v \rangle$ must be computed
numerically using the cut-out distribution function. Details of this
calculation are given in Read (1997) -- there is only a small change
in the ratio $T/|W|$ and the main conclusion of this Section is
unaffected. For the power-law disks, the Ostriker-Peebles criterion is
untrustworthy as it greatly overestimates the amount of energy in the
form of random motion necessary to achieve stability against bar
modes.

%% file: onearmed.tex
\section{One-armed modes}\label{sec:m1chapter}
The observational motivation for studying one-armed modes is that mild
lop-sidedness is seen in many spiral galaxies, with substantial
asymmetries present in some late-type galaxies~\cite{Bald:1980}. This
is evident in both the optical and HI data~\cite{RS:1994,RZ:1995}.
One-armed disturbances differ from many-armed disturbances in one
crucial respect.  It is intuitively clear that one-armed patterns can
move the barycentre from the origin towards the overdense spiral arm.
For an isolated system, acceleration of the barycentre is not
physical. The possibility occurs here because the analysis holds the
equilibrium disk fixed while imposing a perturbation on it. If
one-armed waves grow in a galaxy, the equilibrium disk is displaced to
balance the perturbation so that the barycentre remained fixed.  Our
analysis thus reproduces a famous mistake by Maxwell in his Adams
Prize essay~\cite{Maxwell:1890}. A complete study of the $m=1$ modes
requires that our analysis be modified to permit the barycentre to
move.  Instead, we investigate the stability of the disk to
``na\"{\i}ve'' one-armed disturbances and check {\it a posteriori}
that the barycentre moves only slightly. In this, we are in the good
company of Toomre and Zang, who did exactly this in their analysis of
the disk with a flat rotation curve.

%
\subsection{The Behaviour of the Eigenvalues}
\begin{figure}
	\begin{center}
		\epsfig{file=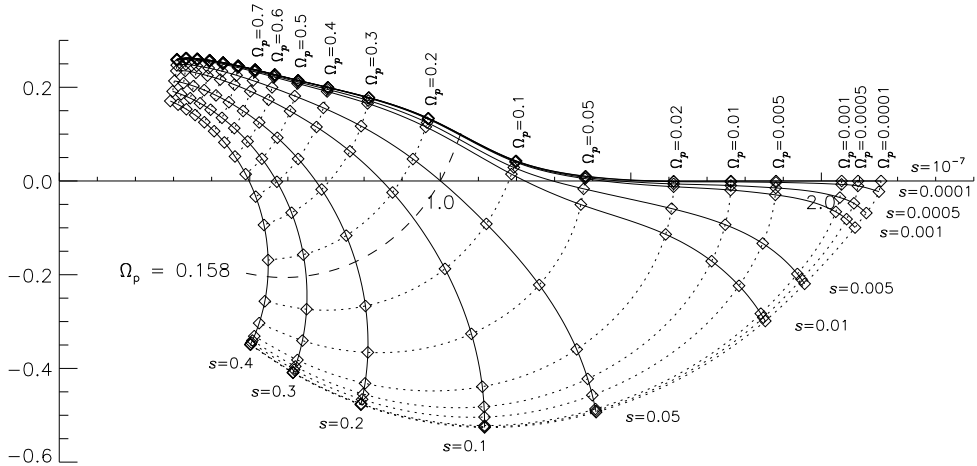,width=0.8\textwidth,height=0.4\textwidth}
		\caption[The dependence of the largest mathematical
		eigenvalue on growth rate and pattern speed, for
		inner-cut-out disks]{The dependence of the largest
		mathematical eigenvalue on growth rate and pattern
		speed for one-armed disturbances in inner cut-out
		disks with $N=2$ and $\Qsing=1$. The horizontal axis
		is $\Re[\lambda]$ and the vertical axis is
		$\Im[\lambda]$.  [In the plot, $\beta=0.25$ ($\sigutil
		= 0.283$ and $\gamma=11.0$).  The solid lines show
		curves of constant $s$ at 11 values of $s$:
		$s={10}^{-7}$, 0.0001, 0.0005, 0.001, 0.005, 0.01,
		0.05, 0.1, 0.2, 0.3, 0.4. The dotted lines show curves
		of constant $\Omegap$ at 20 values of $\Omegap$:
		$\Omegap=0.0001$, 0.0005, 0.001, 0.005, 0.01, 0.02,
		0.05, 0.1, 0.2, 0.3 $\dots$ 1.2, 1.3].
		\label{fig:OmegapLoop113_b0p25_Q1_m1}} \end{center}
\end{figure}

Let us first investigate how the largest mathematical eigenvalue
depends on pattern speed and growth rate for the temperature at which
the self-consistent disk is locally just stable to axisymmetric
disturbances. Fig.~\ref{fig:OmegapLoop113_b0p25_Q1_m1} shows the
eigenvalue curves for a disk with $\beta=0.25$ and $N=2$.
For growing perturbations ($s \gtrsim 0.05$), these curves are
qualitatively similar to those obtained for bisymmetric perturbations
(Fig.~\ref{fig:OmegapLoop113_b0pm25_Q1}). Curves of a given $s$ cross
the real axis at higher values for these $m=1$ plots than for the
earlier $m=2$ plots, indicating that the disk is more prone to
one-armed than two-armed disturbances. For instance, we saw in
Fig.~\ref{fig:OmegapLoop113_b0pm25_Q1} that $\Qsing=1$ was already
sufficient to stabilise the $\beta=0.25$ disk to $m=2$
disturbances. But, Fig.~\ref{fig:OmegapLoop113_b0p25_Q1_m1} indicates
that this disk remains susceptible to $m=1$ disturbances. One such
mode is indicated by the dashed line.  The most striking difference
between the present situation and that for $m=2$ occurs as the growth
rate is reduced to zero. It appears that in the limit of vanishing
growth rate, the marginal eigenvalue curves never cross the real axis.
Eigenvalues for small but non-zero growth rate (e.g. $s=10^{-4},
10^{-3}$) follow the marginal curve for high pattern speeds. As the
pattern speed is reduced, one by one they peel away from the marginal
curve and cross the real axis.  This has dire consquences for the
stability of the disks to one-armed disturbances.  Ordinarily, disks
may be stabilised by raising their temperature, but this does not work
here. As the temperature is increased, the curves with higher growth
rate do indeed shrink towards the origin.  But the curves with very
low growth rate continue to stretch out along the real axis. It
appears that, no matter how high the temperature, a self-consistent
mode exists for sufficiently low growth rate and pattern speed.
Unlike for $m=2$, there is no clear critical temperature
distinguishing stable and unstable disks.  We can only find a
temperature sufficient to stabilise the disk down to
$|\omega|=10^{-3}, 10^{-4}$ or whatever.  Thus, it seems that there is
always at least one mode. The existence of further modes depends on
the behaviour of the sub-dominant eigenvalues.  The sub-dominant
eigenvalues behave in a similar way to the dominant eigenvalue as the
growth rate and pattern speed are varied.

\begin{figure}
	\begin{center}
	\epsfig{file=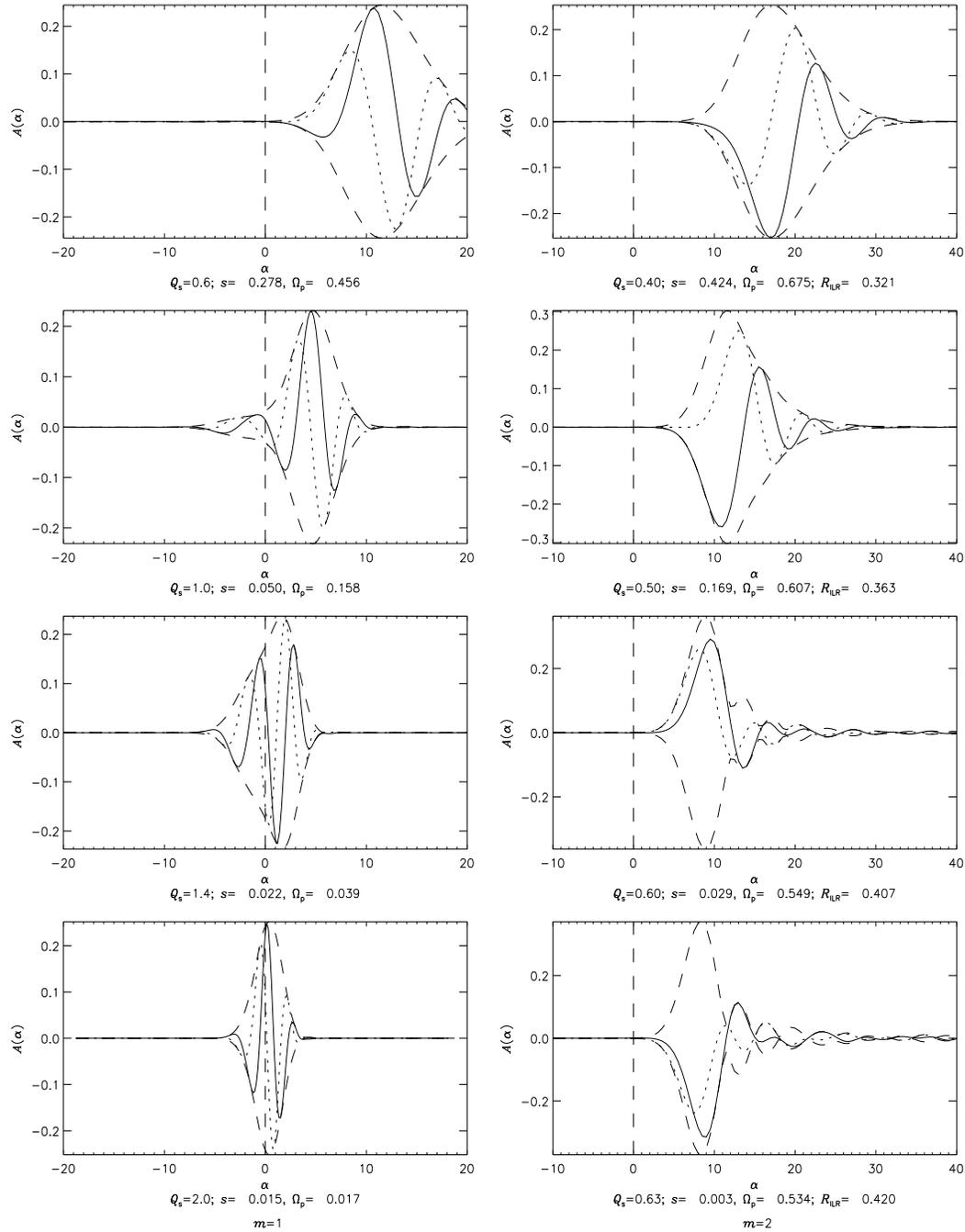,width=0.8\textwidth,height=\textwidth}
	\caption[Density transforms for $m=1$ and $m=2$ modes in disks
	of various temperatures.]{Density transforms for $m=1$ (left,
	$\beta = 0.25$) and $m=2$ (right, $\beta = -0.25$) modes in
	disks of various temperatures. The captions to each plot
	record the value of $\Qsing$, and the growth rate and pattern
	speed of the mode.  For $m=2$, the position of the inner
	Lindblad resonance is also stated.  In each plot, the solid
	curve shows $\Re[A(\alpha)]$, the dotted line
	$\Im[A(\alpha)]$, and the dashed line the envelope
	$\pm|A(\alpha)|$. The dashed vertical line separates leading
	from trailing waves.
\label{fig:m12_LeadingWaves}}
\end{center}
\end{figure}

What is the physical mechanism underlying the instability?  A
distinguishing feature of one-armed perturbations is that they are the
only non-axisymmetric disturbances which have no damping inner
Lindblad resonance.  In the absence of an inner Lindblad resonance,
incoming trailing waves are reflected by the inner cut-out as leading
waves, which are then swing-amplified~\cite{GLB:1965,Toomre:1981}.
For $m=2$, many of the incoming waves are absorbed by the Lindblad
resonance and never return from the central regions.  This point is
illustrated in Fig.~\ref{fig:m12_LeadingWaves}. These plots show the
density transforms of $m=1$ and $m=2$ modes in disks of various
temperatures, chosen so as to obtain modes of broadly similar growth
rates.  The top plot in each figure shows a quickly-growing mode in a
cool disk. Both transforms are dominated by a large trailing
component.  Heating the disk reveals differences between the $m=1$ and
$m=2$ modes. For $m=1$, the presence of a large leading component --
previously hidden by swing-amplified trailing waves -- becomes
apparent. Close to marginal stability, in the bottom plot, the leading
and trailing components are of almost equal amplitude. However, for
$m=2$ modes, no such leading component exists. This indicates that few
leading waves are returning from the central region to fuel the
amplifier. In fact, the physical origin of the $m=2$ modes is quite
different. They occur in such cold disks that they are basically just
Jeans instabilities.
%
%
\subsection{Growing Modes}
\begin{figure}
	\begin{center}
		\epsfig{file=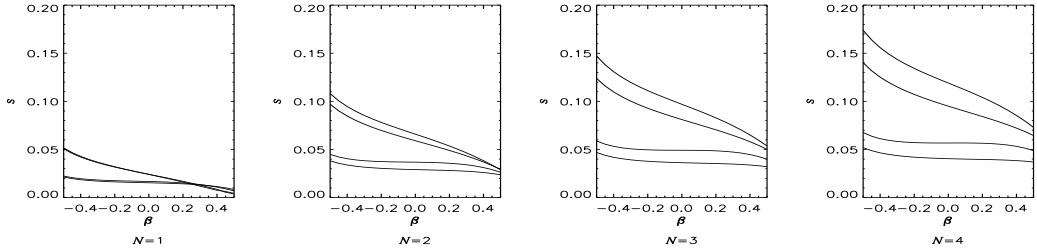,width=0.8\textwidth,height=0.2\textwidth}
		\caption [Growth rates of the four fastest-growing
		modes for inner cut-out disks with $\Qsing=1.0$. ]
		{Growth rates of the four fastest-growing one-armed modes for
		inner cut-out disks with $\Qsing=1.0$. The four plots
		show results for $N=1$, 2, 3, 4.\label{fig:Modes_Q1p0_m1_s}}
		\end{center}
\end{figure}

\begin{figure}
	\begin{center}
		\epsfig{file=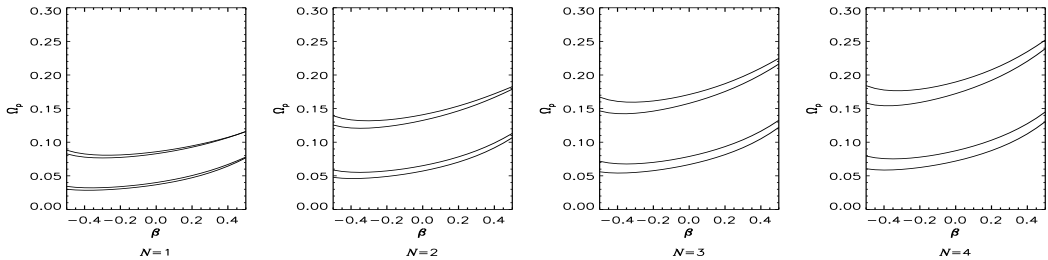,width=0.8\textwidth,height=0.2\textwidth}
		\caption [Pattern speeds of the four fastest-growing
		modes for inner cut-out disks with $\Qsing=1.0$. ]
		{Pattern speeds of the four fastest-growing one-armed modes for
		inner cut-out disks with $\Qsing=1.0$. The four plots
		show results for $N=1$, 2, 3, 4.\label{fig:Modes_Q1p0_m1_Om}}
		\end{center}
\end{figure}

How does the rotation curve and inner cut-out affect the pattern speed
and the growth rate of the one-armed disturbances? To answer this, we
need a way of comparing the stability of different disks.  In the
$m=1$ case, there is no critical temperature above which the disks are
stable. So, instead we employ the growth rate of the fastest-growing
mode in disks with the same temperature as a measure of
stability. More unstable disks admit faster-growing modes.
Fig.~\ref{fig:Modes_Q1p0_m1_s} shows the growth rates obtained for the
four fastest-growing modes in four different inner cut-out disks with
$\Qsing=1$.  Fig.~\ref{fig:Modes_Q1p0_m1_Om} shows the corresponding
pattern speeds. 
Even though the condition $\Qsing=1$ corresponds to larger velocity
dispersions for negative $\beta$, modes still grow faster in disks
with rising rotation curves.  Disks with sharper cut-outs are more
unstable than those where the cut-out is gentle.  For one-armed modes,
there is no inner Lindblad resonance, so even disks with $N=1$ admit
growing modes. However, it appears that a sharper cut-out still
destabilises the disk, presumably because the increased reflection
makes a more efficient feedback circuit for swing amplification.

\begin{figure}
	\begin{center}
		\epsfig{file=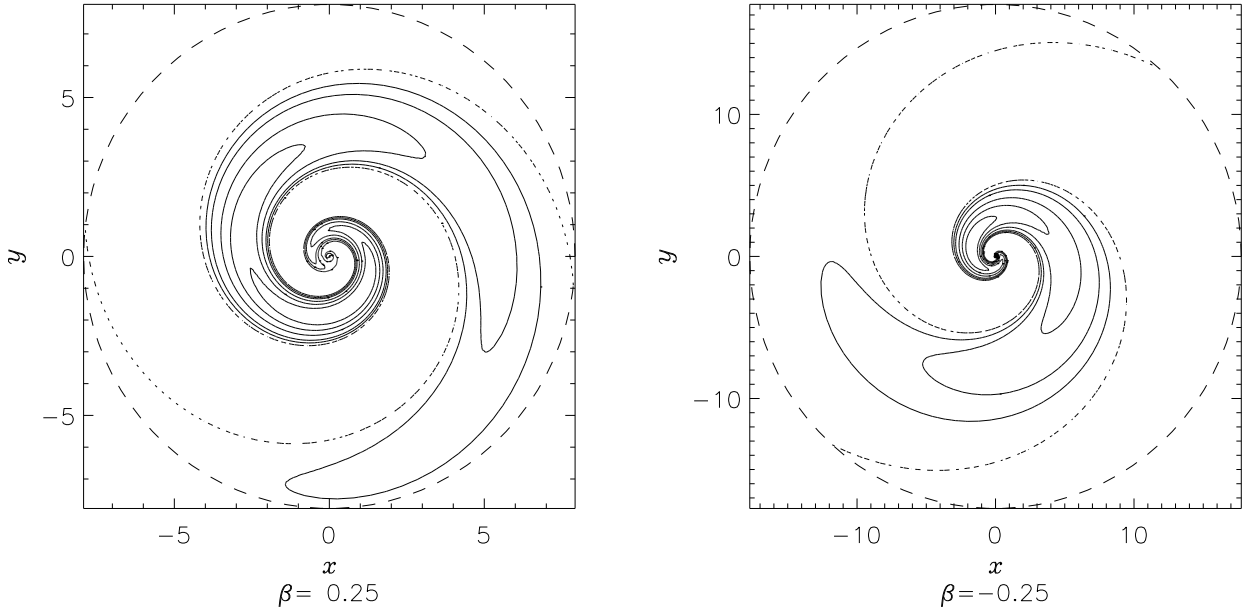,width=\textwidth,height=0.5\textwidth}
		\caption[Density of fastest-growing $\Qsing=1$ mode in
		inner cut-out disks with $N=1$, $\beta=\pm 0.25$.]{
		Density contour plot of the fastest-growing $m=1$ mode
		in inner cut-out disks with $N=1$ and $\Qsing=1$.  The
		solid lines mark 10\%, 20\%, 40\%, 60\% and 80\% of
		the maximum density in this range; the dotted lines
		show the nodes, where the density perturbation is
		zero.  The dashed circle indicates the co-rotation
		radius. [The left-hand plot is for $\beta=0.25$, for
		which the fastest-growing mode has $\Omegap=0.0974$,
		$s=0.0140$, $\alphamu \approx 4.2$.  The right-hand
		plot is for $\beta=-0.25$, for which the
		fastest-growing mode has $\Omegap=0.0807$, $s=0.0343$,
		$\alphamu \approx 2.8$].
		\label{fig:Mode_N1_b0pm25_Q1_m1}} 
\end{center}
\end{figure}

\begin{figure}
	\begin{center}
		\epsfig{file=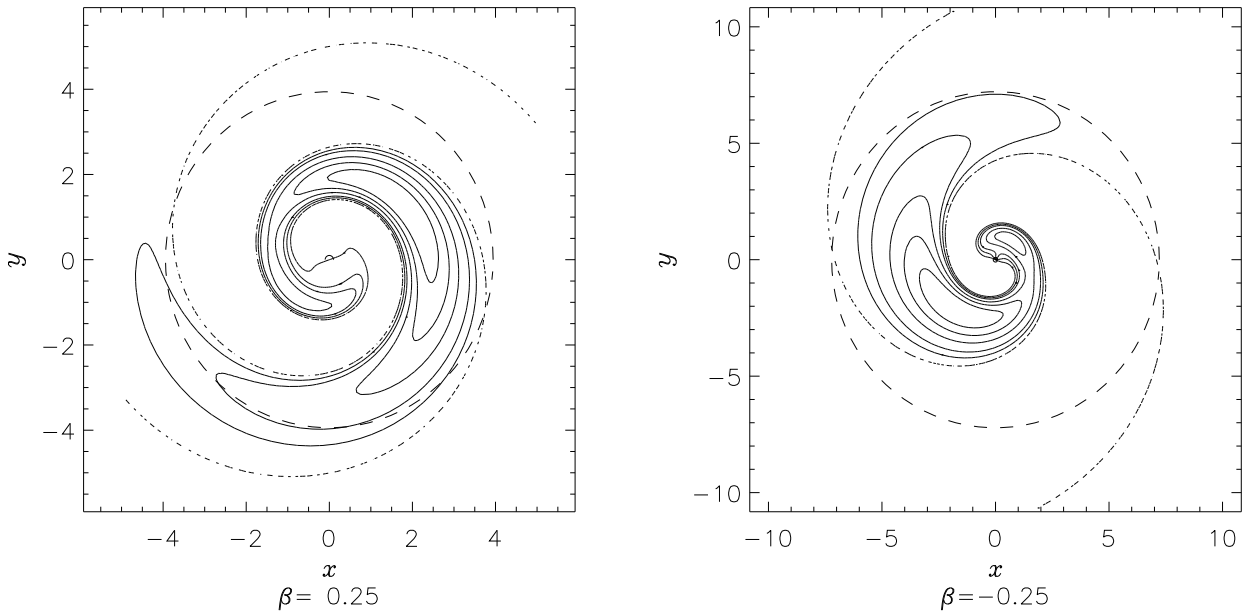,width=\textwidth,height=0.5\textwidth}
		\caption[Density of fastest-growing $\Qsing=1$ mode in
		inner cut-out disks with $N=4$, $\beta=\pm 0.25$.]{
		Density contour plot of the fastest-growing $m=1$ mode
		in inner cut-out disks with $N=4$ and $\Qsing=1$. [The
		left-hand plot is for $\beta=0.25$, for which the
		fastest-growing mode has $\Omegap=0.214$, $s=0.100$,
		$\alphamu \approx 4.6$.  The right-hand plot is for
		$\beta=-0.25$, for which the fastest-growing mode has
		$\Omegap=0.177$, $s=0.139$, $\alphamu \approx 2.8$].
		\label{fig:Mode_N4_b0pm25_Q1_m1}} 
\end{center}
\end{figure}

Figs.~\ref{fig:Mode_N1_b0pm25_Q1_m1} -~\ref{fig:Mode_N4_b0pm25_Q1_m1}
show contour plots of the fastest-growing modes in inner cut-out disks
with $\Qsing=1$. The dominant wavenumber of the modes -- that is, the
wavenumber at which the magnitude of the density transform is
greatest -- is reported in the caption of each figure.
As for the marginally stable bisymmetric modes plotted in
Fig.~\ref{fig:Mode_N2_b0pm25_Qmarg}, the one-armed spirals shown here
are more tightly wound in disks with falling rotation curves than
disks with rising rotation curves.  In accord with the lack of an
inner Lindblad radius, these one-armed modes extend right to the
centre of the disk. Modes in disks with rising rotation curves have
more density concentrated close to the centre than those in disks with
falling rotation curves.  When discussing the distance to which these
modes extend out in the disk, it is useful to distinguish between the
distance in physical space, and the distance in terms of the
co-rotation radius.  Modes in disks with low velocity dispersions
generally extend further beyond the co-rotation radius than modes in
warmer disks. However, modes in cooler disks also rotate more quickly,
so their co-rotation and outer Lindblad radii are smaller. Thus the
actual extent in space of a mode is generally smaller in cool disks
than in warm disks.  Bearing these temperature effects in mind, there
is little difference in terms of spatial extent between one- and
two-armed modes.  The one-armed modes shown in
Figs.~\ref{fig:Mode_N1_b0pm25_Q1_m1} - ~\ref{fig:Mode_N4_b0pm25_Q1_m1}
extend further in space, but less far beyond the co-rotation radius,
than the two-armed modes plotted in
Figs.~\ref{fig:Mode_N2_b0pm25_Qmarg}. This reflects the higher
temperature of the disks in which the $m=1$ modes are plotted.

%
\subsection{The Position of the Barycentre}
One-armed modes are unique in that they are capable of displacing the
barycentre of the disk away from the origin.  Visual inspection of
Figs.~\ref{fig:Mode_N1_b0pm25_Q1_m1} and
~\ref{fig:Mode_N4_b0pm25_Q1_m1} suggests that in these modes the
barycentre is still close to the origin.  The spiral arm wraps around
the origin, with more density in the parts of the arm closer to the
origin. It is difficult to guess by eye even the direction of any
displacement of the barycentre from the origin.  Intriguingly, the
density distribution in one of the cases shown has two separate peaks,
one on either side of the origin. This further suggests a balancing of
the density on either side of the origin.

\begin{figure}
	\begin{center}
		\epsfig{file=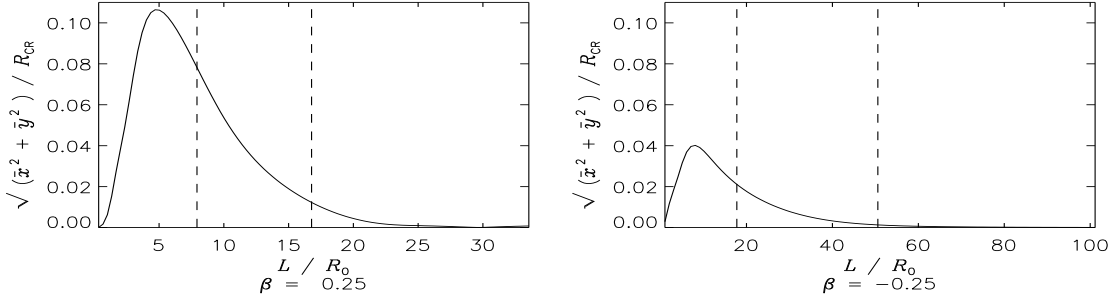,width=0.85\textwidth,height=0.25\textwidth}
		\caption [The displacement of the barycentre from the
		origin, for $N=1$ disks with $\beta = \pm 0.25$] {The
		displacement of the barycentre from the origin for
		$N=1$ disks with $\beta = \pm 0.25$. The displacement
		is plotted in units of the co-rotation radius. We have
		set $\Sigmap= \Sigma_0$ to make the point that the
		barycentre does not move to second order.  The dashed
		vertical lines mark the position of the co-rotation
		and outer Lindblad radii.
\label{fig:CentreOfMass_N1_Q1}}
	\end{center}
\end{figure}

In general, the coordinates of the barycentre of the part of the disk
within a radius $L$ are given by
\begin{xalignat}{2}
	M(L) \bar{x} & = \int_0^L dR \int_0^{2\pi} d\theta
	\Sigma(R,\theta) R^2 \cos \theta , &\quad M(L) \bar{y} & =
	\int_0^L dR \int_0^{2\pi} d\theta \Sigma(R,\theta) R^2 \sin
	\theta , \label{eq:xbarybar}
\end{xalignat}
where $M(L)$ is the mass contained within the radius $L$.  Each
logarithmic spiral component makes equal positive and negative
contributions at every radius, so the mass $M(L)$ contained within a
radius $L$ is just the integral of the equilibrium density.
Conversely, as the equilibrium density is axisymmetric, it does
not shift the barycentre from the origin. So, the expression
~\eqref{eq:xbarybar} for the $x$-coordinate of the barycentre becomes
(using eqs. (44-45) of Paper I)
\begin{equation}
	\frac{4\pi }{1-\beta} 
	\left( \frac{L}{R_0} \right)^{1-\beta}
	\frac{\Sigma_0}{\Sigmap}
	\bar{x} 
	= 
	\frac{ e ^{(s - im \Omegap )t}}{R_0^{1/2+i\alpha}} 
	\int_{-\infty}^{+\infty} d\alpha A( \alpha ) 
	\int_0^L dR \, R^{1/2+i\alpha}
	\int_0^{2\pi} d\theta 
	\left\{  e ^{ i(m+1)\theta   } + e ^{ i(m-1)\theta } \right\}
\end{equation}
where we must take the real part to obtain the physical co-ordinates.
For $m=1$, we obtain
\begin{equation}
	\bar{x} 
	= 
	\frac{1-\beta} {2 }
	\frac{\Sigmap}{\Sigma_0}
	R_0 e ^{st }
	\left( \frac{L}{R_0} \right)^{1/2+\beta}
	\Re	
\left\{	
	e^{ - i \Omegap t }
	\int_{-\infty}^{+\infty} d\alpha A( \alpha ) 
	\left( \frac{L}{R_0} \right)^{i\alpha}
\right\},
	\label{eq:xbar}
\end{equation}
\begin{equation}
	\bar{y} 
	= 
	-\frac{1-\beta} {2 }
	\frac{\Sigmap}{\Sigma_0}
	R_0 e ^{st}
	\left( \frac{L}{R_0} \right)^{1/2+\beta}
	\Im
\left\{	
	e^{ - i \Omegap t }
	\int_{-\infty}^{+\infty} d\alpha A( \alpha ) 
	\left( \frac{L}{R_0} \right)^{i\alpha}
\right\}.
	\label{eq:ybar}
\end{equation}
If the integral over $\alpha$ is non-zero, the barycentre is displaced
from the origin.  As $L$ is taken to infinity, the coordinates of the
barycentre tend to infinity for $\beta > -\fr12$. If we consider a
large but finite value of $L$, eqs.~\eqref{eq:xbar} and
~\eqref{eq:ybar} show that the barycentre gradually spirals
outwards. The distance of the barycentre from the origin is given by
\begin{equation}
	\sqrt{\bar{x}^2+\bar{y}^2}
	= 
	\frac{1-\beta} {2 }
	\frac{\Sigmap}{\Sigma_0}
	R_0 e ^{st }
	\left( \frac{L}{R_0} \right)^{1/2+\beta}
\left|
	\int_{-\infty}^{+\infty} d\alpha A( \alpha ) 
	\left( \frac{L}{R_0} \right)^{i\alpha}
\right|.
	\label{eq:modxbarybar}
\end{equation}
Fig.~\ref{fig:CentreOfMass_N1_Q1} shows the distance from the origin
of the barycentre of the part of the disk within a radius $L$ in units
of the co-rotation radius, i.e., $\sqrt{\bar{x}^2+\bar{y}^2} /
R_{\text{CR}}$.  The co-rotation radius roughly delimits the edge of
the spiral mode, so this quantity describes how far the barycentre is
displaced from the origin relative to the overall size of the
mode. The dashed vertical lines in Fig.~\ref{fig:CentreOfMass_N1_Q1}
indicate the positions of the co-rotation and outer Lindblad
resonances.
For small values of $L$, the displacement of the barycentre is
approximately linear. This is as expected -- if we consider only the
very inner region of the disk, the emerging spiral arm is
predominantly on one side of the origin. But as we move outwards, the
contributions of the spiral arm from opposite sides of the origin
begin to cancel out.  For values of $L$ beyond the outer Lindblad
resonance, the part of the disk under consideration is now large
enough to include the entire spiral mode. The forces pulling the
barycentre away from the origin in different directions cancel almost
exactly, so that the barycentre of the whole disk remains close to the
origin.  The fast oscillation of the term $\exp(i\alpha \ln L)$ is
probably the reason for the cancellation. For typical modes,
$A(\alpha)$ is a smooth function of $\alpha$ which tends to zero as
$\alpha \rightarrow \pm \infty$. As $L$ is increased, the term
$\exp(i\alpha \ln L)$ must eventually vary much faster with $\alpha$
than does $A(\alpha)$ (unless $A(\alpha)$ is a delta-function or some
other strongly peaked entity). Thus the integration over wavenumber in
eq.~\eqref{eq:modxbarybar} adds up many positive and negative
contributions of nearly equal magnitude. These roughly cancel,
ensuring that the integral almost vanishes.  At least in the linear
r\'egime, it appears that self-consistent modes do not shift the
barycentre of the disk substantially.

%% file: triskele.tex
\section{The Higher Angular Harmonics}\label{sec:m34}
This section briefly examines modes with $m=3$ and 4. These are dubbed
respectively triskele and tetraskele modes, after the three- and
four-legged devices common in early European art. The observational
motivation for studying these modes comes from Fourier decompositions
of the spiral patterns of real galaxies, which indicate significant
contributions from $m=3$ and $m=4$ harmonics~\cite{Elmegreens:1985}. 
The same result is suggested by analysis of N-body bars, which tend 
to be more rectangular than purely elliptical~\cite{SS:1987}.
%
%
\subsection{A Numerical Delicacy}
Modes with $ m \ge 2$ present an additional numerical delicacy. The
expressions for the angular momentum function derived in Appendix C of
Paper I involve division by the combination $\lkapmom$.  When $\lkapmom$
vanishes, the integral still exists in a Cauchy principal value sense.
In the modes examined so far, this problem has not arisen.  For $m=0$,
$\lkapmom$ vanishes at the $l=0$ radial harmonic; but in this case the
angular momentum function $F_{00}$ is simply zero. For $m=1$ and
$m=2$, $\lkapmom$ never vanishes. But for higher angular harmonics,
such as $m=3$ and $m=4$, the term $\lkapmom$ may vanish at some
eccentric velocity.

What does it mean physically when $\lkapmom$ vanishes? The time the
star takes to perform $l$ complete revolutions, $2\pi |l|/\Omega$, is
then identical to the time it takes to perform $m$ radial
oscillations, $2\pi m /\kappa$. Thus the star's orbit closes in an
inertial frame.  Closed orbits such as these do not occur in all
disks. For $\lkapmom$ to vanish, then in terms of the auxiliary
integral (defined in eq. (18) of Paper I), $\auxint_2 (\Util) = - 2\pi
l/m$.  In section 2 of Paper I, we saw that $\auxint_2 (\Util)$
declines monotonically from $2\pi / \sqrt{2-\beta}$ at $\Util=0$ to
$\pi$ at $\Util = \infty$. Thus $\lkapmom$ will vanish for some finite
$\Util$ if there is an integer $l$ which satisfies
\be
	\half < - \frac{l}{m} < \frac{1}{\sqrt{2-\beta}}.
	\label{eq:lkapmomvanish}
\end{equation}
Whether or not this occurs depends on the values of $m$ and $\beta$.
For $m=3$, $\lkapmom$ vanishes at $l=-2$ for $\beta > -0.25$, while
for $\beta < -0.25$ it never vanishes. For $\beta=-0.25$, $\lkapmom$
vanishes at $\Util=0$, meaning that the relevant orbits are circular.
For $m=4$, $\lkapmom$ vanishes at $l=-3$ for $\beta > 2/9$, while for
$\beta < 2/9$ it never vanishes.  Even when $\lkapmom$ vanishes, the
mathematical eigenvalue can still be evaluated
accurately~\cite{Read:1997}.  The Gauss-Laguerre quadrature used for
the integration over eccentric velocity successfully interpolates
across the troublesome value of $\Util$.
%
%
\subsection{Triskele or Three-Armed Modes\label{sec:threearmed}}
\subsubsection{Global Stability of the Cut-Out Disks}
\begin{figure}
	\begin{center}
		\epsfig{file=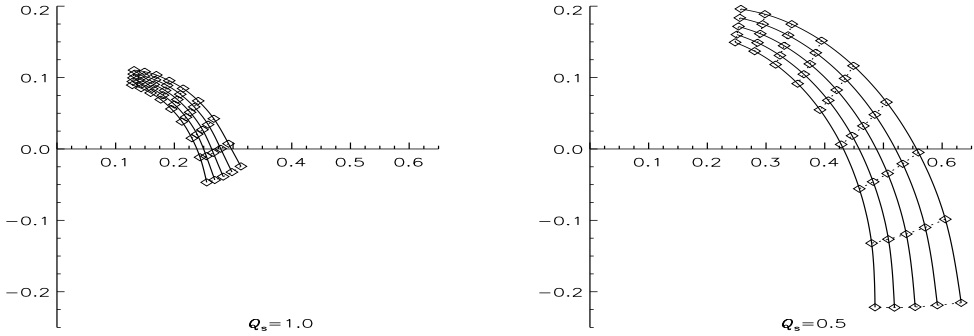,width=0.8\textwidth,height=0.25\textwidth}
		\caption [The dependence of the largest eigenvalue on
		growth rate and pattern speed, for an inner-cut-out
		disk with $\beta=0.25$] {The dependence of the largest
		eigenvalue on growth rate and pattern speed for $m=3$
		disturbances in cut-out disks with $N=2$ and
		$\beta=0.25$. Note that the curves cross the real axis
		well to the left of the point ($1,0$) required for
		self-consistent modes. [The left-hand plot is for
		$\Qsing=1.0$, i.e. $\sigutil = 0.283$ and
		$\gamma=11.0$. The right-hand plot is for
		$\Qsing=0.5$, i.e. $\sigutil=0.141$, $\gamma =
		48.5$. The solid lines show curves of constant $s$ at
		intervals of 0.1 from $s=0^+$ (right-most curve) to
		$s=0.4$ (left-most curve). The dotted lines show
		curves of constant $\Omegap$ at intervals of 0.1 from
		$\Omegap=0.5$ (lowest curve) to $\Omegap=1.3$ (highest
		curve)]. \label{fig:OmegapLoop113_b0p25_Q_m3}}
		\end{center}
\end{figure}

Fig.~\ref{fig:OmegapLoop113_b0p25_Q_m3} shows the dependence of the
largest eigenvalue on growth rate and pattern speed for
$\beta=0.25$. The left panel shows the results for $\Qsing=1.0$, the
right panel for $\Qsing=0.5$.
This plot already suggests that the power-law disks are very stable to
$m=3$ perturbations. Even when the temperature is as low as
$\Qsing=0.5$, the eigenvalue curves intersect the real axis at values
considerably less than unity.  For disks with $N=2$, $\Qsing$ has to
be lower than 0.4 before even the $\beta=-0.5$ disk becomes unstable
to $m=3$ perturbations. For disks with falling rotation curves, even
lower temperatures are necessary. At these very low temperatures, the
convergence of the numerical method is somewhat untrustworthy.  This
is mainly because the modes are more tightly wound at low
temperatures. Thus, the density transform tends to be oscillatory and
peaked at high wavenumber, so that large and finely-meshed grids are
required.

\begin{figure}
	\begin{center}
		\epsfig{file=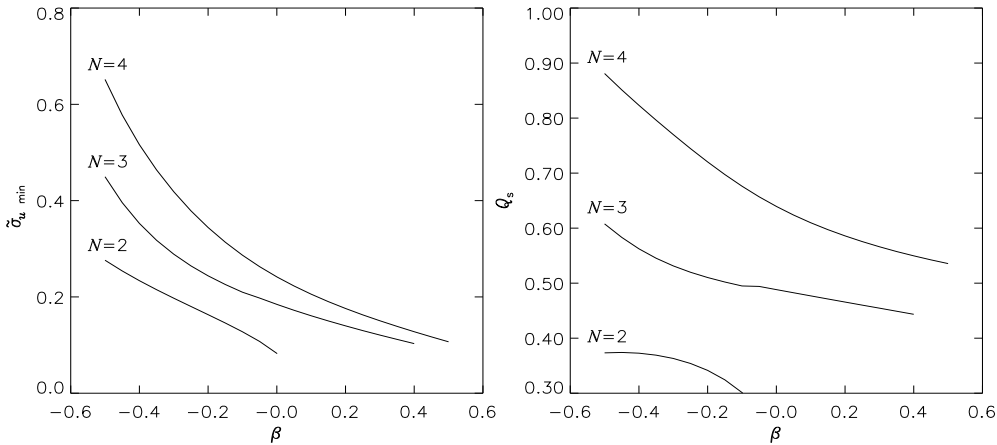,width=0.75\textwidth,height=0.35\textwidth}
		\caption[Minimum temperature plotted against $\beta$,
		for cut-out disks] {Minimum temperature for stability
		against $m=3$ modes plotted against rotation curve
		index $\beta$ for disks with various cut-out
		functions. The left-hand plot shows the minimum
		velocity dispersion $\sigutil$; the right-hand one the
		same data presented in terms of the stability
		parameter $\Qsing$. The solid lines indicate the
		results for inner cut-out disks with $N=2$, 3,
		4.
                \label{fig:m3margsiggam}} \end{center}
\end{figure}

Fig.~\ref{fig:m3margsiggam} shows the temperature of the marginal
modes, presented in terms of the velocity dispersion $\sigutil$ and
stability parameter $\Qsing$.  The $N=2$ results are probably not
trustworthy to better than 2 s.f., but are included since they give at
least an indication of the very low temperatures required.  For $N=3$
and $N=4$, the temperature is higher and hence the convergence is
better.
The rotation curve and inner cut-out affect the stability to
three-armed perturbations in the same way as they affect the $m=2$
stability. Disks where the centre has been cut out more sharply are
more unstable than those where it has been removed relatively
gently. Power-law disks with rising rotation curves are less securely
stable than those with falling rotation curves.
%
%
\subsubsection{Neutral Modes}
\begin{figure}
	\begin{center}
		\epsfig{file=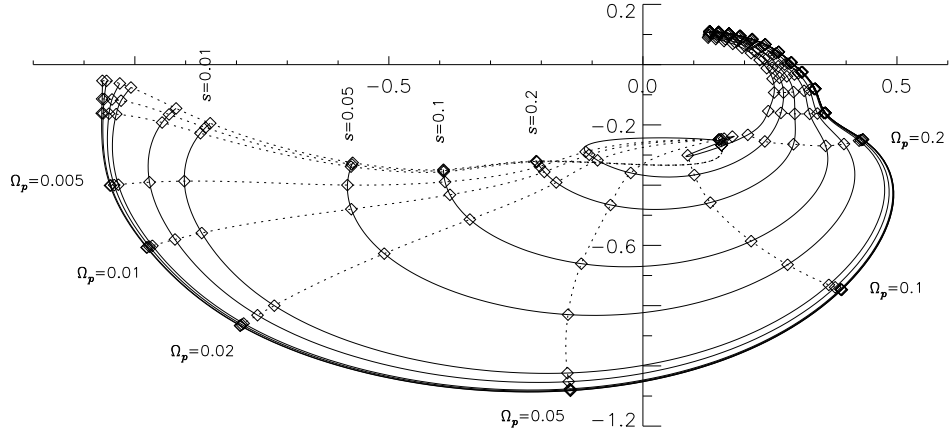,width=0.8\textwidth,height=0.4\textwidth}
		\caption [The dependence of the largest eigenvalue on
		growth rate and pattern speed, for an inner-cut-out
		disk with $\beta=0.25$] {The dependence of the largest
		mathematical eigenvalue on growth rate and pattern
		speed for $m=3$ disturbances in a disk with $N=2$,
		$\Qsing =1$ and $\beta=0.25$.  The horizontal axis is
		$\Re[\lambda]$ and the vertical axis is
		$\Im[\lambda]$.  [The solid lines show curves of
		constant $s$ at 11 values of $s$: $s={10}^{-7}$,
		0.0001, 0.0005, 0.001, 0.005, 0.01, 0.05, 0.1, 0.2,
		0.3, 0.4. The dotted lines show curves of constant
		$\Omegap$ at 20 values of $\Omegap$: $\Omegap=0.0001$,
		0.0005, 0.001, 0.005, 0.01, 0.02, 0.05, 0.1, 0.2, 0.3
		$\dots$ 1.2, 1.3].
		\label{fig:OmegapLoop119_b0p25_Q1_m3}} \end{center}
\end{figure}

\begin{figure}
	\begin{center}
		\epsfig{file=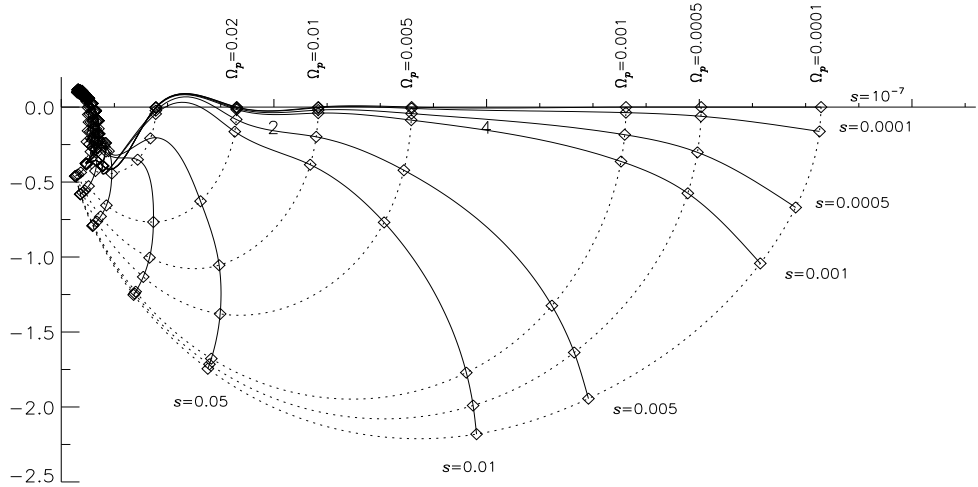,width=0.8\textwidth,height=0.4\textwidth}
		\caption [The dependence of the largest eigenvalue on
		growth rate and pattern speed, for an inner-cut-out
		disk with $\beta=-0.25$] {The dependence of the
		largest mathematical eigenvalue on growth rate and
		pattern speed for $m=3$ disturbances in a disk with
		$N=2$, $\Qsing =1$ and $\beta=-0.25$.  The horizontal
		axis is $\Re[\lambda]$ and the vertical axis is
		$\Im[\lambda]$.  [The solid lines and dotted lines are
		as in Fig.~\ref{fig:OmegapLoop119_b0p25_Q1_m3}].
		\label{fig:OmegapLoop119_b0m25_Q1_m3}} \end{center}
\end{figure}

So far, it appears as if the stability of the power-law disks to $m=3$
perturbations is much like that to $m=2$ perturbations, the only
difference being that even lower temperatures are required before
growing modes can be excited. This conclusion seems to be reasonable
enough in the light of eigenvalue curves such as
Fig.~\ref{fig:OmegapLoop113_b0p25_Q_m3}.  However, a different picture
emerges when we pursue the mathematical eigenvalues down to
vanishingly low growth rates and pattern speeds.
Figs.~\ref{fig:OmegapLoop119_b0p25_Q1_m3} and
~\ref{fig:OmegapLoop119_b0m25_Q1_m3} show the dominant eigenvalue
curves as $s$ and $\Omegap$ tend to zero, for $\Qsing=1$ and
$\beta=\pm0.25$.
For $\beta=0.25$, the behaviour is similar to that observed for $m=2$
perturbations.  As the pattern speed is decreased, the eigenvalue
curves continue to arc clockwise around the origin, tending to a a
constant value (for given growth rate) in the limit $\Omegap
\rightarrow 0$.  They appear to be tending towards the negative real
axis in the twin limits $s \rightarrow 0$, $\Omegap \rightarrow 0$.
However, for $\beta=-0.25$ the behaviour is strikingly different.
Fig.~\ref{fig:OmegapLoop119_b0m25_Q1_m3} is much closer to the $m=1$
eigenvalue curves shown in
fig.~\ref{fig:OmegapLoop113_b0p25_Q1_m1}. It indicates that that modes
with very low pattern speeds and growth rates are possible even at the
high temperature of $\Qsing=1.0$.  It appears that there is a dramatic
change in the behaviour of the eigenvalue curves between $\beta=+0.25$
and $\beta=-0.25$.

\begin{figure}
	\begin{center}
		\epsfig{file=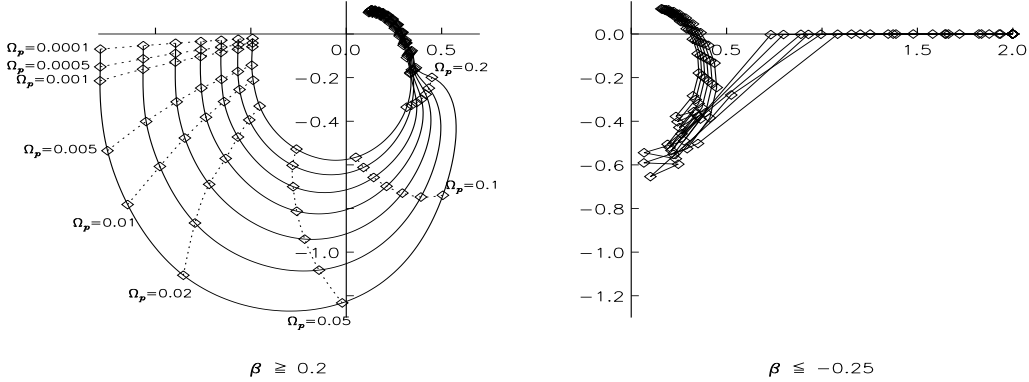,width=0.8\textwidth,height=0.3\textwidth}
		\caption [Marginal eigenvalue curves for inner cut-out
		disks with $\Qsing=1.0$ and different values of
		$\beta$] {Marginal eigenvalue curves for $m=3$
		disturbances in inner cut-out disks with $N=2$,
		$\Qsing=1.0$ and various values of $\beta$.  The left
		panel shows these marginal eigenvalue curves for
		positive values of $\beta \geq 0.2$, namely
		$\beta=0.2$, 0.25, 0.3 $\dots$ 0.45, 0.5. The right
		panel shows the curves for values of $\beta \leq
		-0.25$, namely $\beta=-0.25$, -0.3, -0.35 $\dots$
		-0.45, -0.5. For intermediate values of $\beta$, the
		marginal eigenvalue curves are tangled, and are not
		shown here. Their behaviour is closer to the curves
		with $\beta \geq 0.2$ than to those with $\beta \leq
		-0.25$.  [Each plot shows the largest mathematical
		eigenvalues for vanishing growth rate, $s=10^{-7}$,
		and 20 values of the pattern speed: $\Omegap =
		0.0001$, 0.0005, 0.001, 0.005, 0.01, 0.02, 0.05, 0.1,
		0.2, 0.3, $\dots$ 1.2, 1.3.]
\label{fig:MarginalManyBeta_Q1_m3}}
		\end{center}
\end{figure}

It has proved hard to investigate the precise value of $\beta$ at
which such neutral modes become possible, since for values of $\beta$
slightly greater than $-0.25$, the convergence becomes problematic.
Fig.~\ref{fig:MarginalManyBeta_Q1_m3} presents evidence which is at
least consistent with the idea that $\beta=-0.25$ is the delimiting
value. The left-hand plot of Fig.~\ref{fig:MarginalManyBeta_Q1_m3}
shows marginal eigenvalue curves for seven positive values of
$\beta$. None admit neutral modes. The right-hand plot shows marginal
eigenvalue curves for six values of $\beta \leq -0.25$. All admit
neutral modes.  Fig.~\ref{fig:MarginalManyBeta_Q1_m3} is for disks at
$\Qsing=1$. Similar curves were obtained at $\Qsing=0.3$ (not
shown). The evidence suggests a link between the absence of neutral
modes and the presence of orbits which close in an inertial frame. For
$\beta > -0.25$, it appears that the presence of three-lobed closed
orbits prevents the formation of these neutral modes, even at low
temperatures.  For $\beta < -0.25$, neutral modes persist even at high
temperatures.
%

%% file: tetraskele.tex
\subsection{Tetraskele or Four-Armed Modes\label{sec:fourarmed}}
\begin{figure}
	\begin{center}
		\epsfig{file=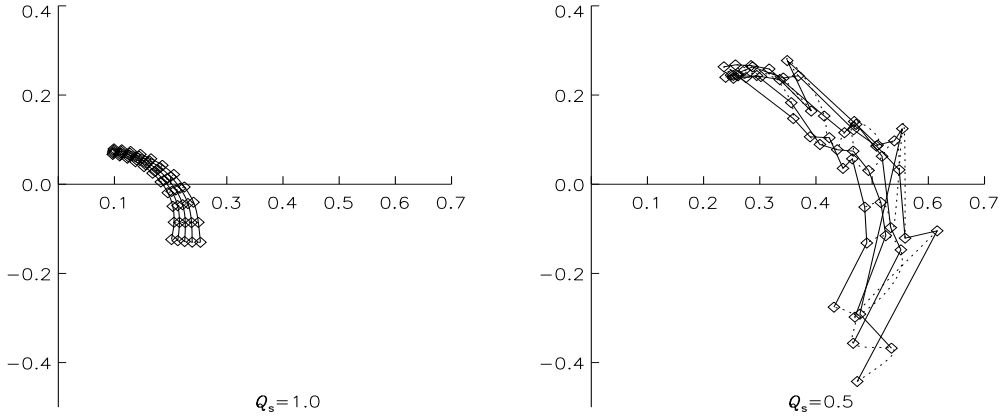,width=0.8\textwidth,height=0.35\textwidth}
		\caption[The dependence of the largest eigenvalue on
		growth rate and pattern speed, for an $m=4$
		perturbation in an inner-cut-out disk with
		$\beta=0.25$]{The dependence of the largest eigenvalue
		on growth rate and pattern speed for $m=4$
		disturbances in $N=2$ and $\beta=0.25$ disks. The left
		panel is for $\Qsing=1.0$, the right panel for
		$\Qsing=0.5$. [The solid lines show curves of constant
		$s$ at intervals of 0.1 from $s=0^+$ (right-most
		curve) to $s=0.4$ (left-most curve). The dotted lines
		show curves of constant $\Omegap$ at intervals of 0.1
		from $\Omegap=0.3$ (lowest curve) to $\Omegap=1.3$
		(highest curve)].
                \label{fig:OmegapLoop113_b0p25_Q1_m4}}
		\end{center}
\end{figure}

\begin{figure}
	\begin{center}
		\epsfig{file=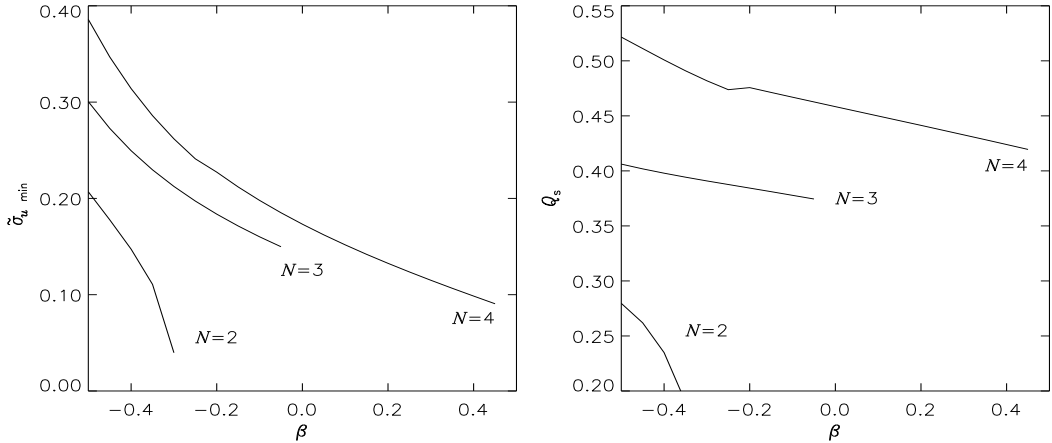,width=0.8\textwidth,height=0.35\textwidth}
		\caption [Minimum temperature plotted against $\beta$,
		for cut-out disks] {Minimum temperature for stability
		against $m=4$ modes plotted against the rotation curve
		index $\beta$ for cut-out disks. The left-hand plot
		shows the minimum velocity dispersion $\sigutil$; the
		right-hand one the same data presented in terms of the
		stability parameter $\Qsing$. The solid lines indicate
		the results for inner cut-out disks with $N=1$, 2, 3,
		4.
                \label{fig:m4margsiggam}}
                \end{center}
\end{figure}

Let us begin our study by examining the eigenvalue curves at moderate
pattern speeds. Fig.~\ref{fig:OmegapLoop113_b0p25_Q1_m4} shows sample
curves for $\beta=0.25$. As before, the left-hand plot is for
$\Qsing=1$, and the right-hand plot for $\Qsing=0.5$. In the latter
case, the eigenvalue curves are ragged, as different eigenvalues
become dominant.
It is immediately apparent that the power-law disks are very stable to
$m=4$ perturbations.  Fig.~\ref{fig:m4margsiggam} shows how low the
temperature must be, in terms of the velocity dispersion $\sigutil$
and the stability parameter $\Qsing$, before the marginal eigenvalue
curve intersects the real axis at unity. Even lower temperatures are
required than for $m=3$ (c.f., Fig.~\ref{fig:m3margsiggam}).  At these
low temperatures the convergence is delicate. The values shown for
$\sigutilmin$ and $\Qsing$ in Fig.~\ref{fig:m4margsiggam} are probably
accurate to only 2 s.f.
Once again, disks where the centre has been cut out more sharply are
more unstable than those where it has been removed relatively
gently. Disks with rising rotation curves are more unstable than those
with falling rotation curves, both in terms of $\sigutilmin$, the
absolute amount of velocity dispersion required, and in terms of
$\Qsing$, the ratio of the velocity dispersion needed relative to that
necessary for axisymmetric stability.

In the absence of closed four-lobed orbits, neutral $m=4$ modes may be
possible. However, these are more easily vanquished by temperature
than the neutral $m=3$ modes. In disks with $\beta \geq 2/9$, the
four-lobed orbits closed in the inertial frame seemingly preclude the
existence of neutral modes with $\Omegap =0$ even at low
temperatures. However, disks with $\beta$ only slightly less than the
dividing possess very slowly growing modes that persist to quite high
temperatures ($\Qsing=1$).  So, their marginal eigenvalue curves end
up on the positive real axis.

%% file: selfcondisk.tex
\section{The Self-Consistent Disk}\label{sec:SelfCon}
Our attention now turns to the self-consistent power-law disk. Here,
there is no cut-out function removing part of the density, and the
potential experienced by stars in the disk is entirely due to their
own gravity. Section~\ref{sec:growing} considers whether growing modes
are possible in the self-consistent disk, while
Section~\ref{sec:neutral} demonstrates the existence of neutral modes.
%
\subsection{Growing Modes\label{sec:growing}}
The self-consistent disk is perfectly scale-free and so possesses no
intrinsic length- or time-scales.  This leads to some rather strange
effects.  As originally pointed out by Kalnajs in 1974 (see Zang's
1976 thesis), all dependence on growth rate and pattern speed can now
be factored out of the integral equation. If a normal mode solution to
the integral equation is found for some growth rate and pattern speed,
it can be scaled to provide solutions at all growth rates and pattern
speeds.  This is proved for the family of self-consistent, power-law
disks in Appendix~\ref{sec:symmetry}.  We shall make the same point
here by dimensional analysis.

Let us recall (see e.g., Section 2 of Paper I) that the equilibrium of the
self-consistent disk contains three parameters -- namely, the
reference radius $R_0$, the reference density $\Sigma_0$ and reference
velocity $v_\beta$. They are related by $v_\beta^2 = 2 \pi G \Sigma_0
R_0 L(\beta)$.  At first sight, it thus appears that we have two
``degrees of freedom''. For example, we could choose $\Sigma_0$ and
$R_0$ with the consequence that $v_\beta$ is then fixed. If this were
the case, $R_0$ would provide a length-scale and $R_0/v_\beta$ a
time-scale. However, one of these ``degrees of freedom'' is
illusory. The surface density is $\Sigma = \Sigma_0 R_0^{1+\beta}
R^{-1-\beta}$, so we cannot tell the difference between disks with the
same $\Sigma_0 R_0^{1+\beta}$. We are therefore free only to choose
$\Sigma_0 R_0^{1+\beta}$, which then determines $v_\beta^2
R_0^\beta$. The disk is characterised by only two quantities, one with
dimensions $[M][L]^{\beta-1}$, and the other with
$[L]^{4+\beta}[T]^{-2}$. It is apparent that these cannot be combined
to yield a quantity with dimensions of length or of
time~\cite{LBL:1993}.  Of course, carving out the centre of the disk
breaks the self-similarity. The radius $R=R_0$ at which the inner
cut-out is applied defines a length-scale. A time-scale is provided by
the period of a circular orbit at this radius. The Fredholm integral
equation then depends on the frequency $\omtil$. We are free to adjust
$\omtil$ so as to obtain non-trivial solutions, or modes.  The
self-similarity of the scale-free disk removes this freedom. The only
adjustable parameter is the temperature. If the disk is hot, no modes
exist.  Intuitively, one expects that as the velocity dispersion is
diminished below a critical value, modes become possible. The symmetry
of the disk prevents it from selecting modes with a special frequency
(unless it is zero).  Thus, the self-consistent disk may possess
isolated neutral modes. As regards growing modes, it may admit
everything (a two-dimensional continuum with all possible frequencies)
or nothing.

The Fredholm integral equation~\eqref{eq:inteqnsing} is hard to solve
numerically for the self-consistent disk, since the kernel is highly
singular.  It immediately seems unlikely that growing,
non-axisymmetric modes could exist in the self-consistent disk. All
the quantities in the Fredholm integral equation~\eqref{eq:inteqnsing}
are complex, so there are effectively two equations: $\Re(\lambda)=1$
and $\Im(\lambda)=0$ must be simultaneously satisfied.  In the cut-out
case, there are two free parameters: the frequency $\omega$ and
temperature $\gamma$. In the self-consistent case, there is only one
free parameter: the temperature $\gamma$. It seems unlikely that both
the real and the imaginary parts of the integral equation could
simultaneously be satisfied at any temperature $\gamma$.  It seems
even less likely that this can occur over a range of temperatures
$\gamma$, as must happen if the growing modes affect all disks below a
critical temperature.  In the light of our analysis of the cut-out
disks, there is one further piece of evidence that can be adduced.
Non-axisymmetric modes in disks are principally confined to the region
between the inner and outer Lindblad radii. As the pattern speed
becomes vanishingly small, the inner Lindblad radius moves well beyond
the inner cut-out radius. Under such circumstances, the response of
the inner cut-out disk must approach that of the self-consistent disk
(c.f., Zang 1976). If a continuum of modes exists in the
self-consistent disk, its presence should be sensed by the cut-out
disks as the pattern speed is brought to zero. We would expect the
eigenvalues to draw together and become virtually independent of
growth rate as the pattern speed approaches zero. There is no evidence
that this happens. Here, the reader is urged to look back at graphs
such as Fig.~\ref{fig:OmegapLoop113_b0p25_Q1_m1} ($m=1$), or
Figs.~\ref{fig:OmegapLoop119_b0p25_Q1_m3} -
~\ref{fig:OmegapLoop119_b0m25_Q1_m3} ($m=3$).  Altogether, the weight
of the evidence seems to us to suggest that there is no continuum and
that the self-consistent disks admit no growing non-axisymmetric modes
at all.

The case of the axisymmetric modes is different. The work in Section
3.5 of Paper I shows that the kernel of the integral equation is
Hermitian and the eigenvalues are real and positive for the $m = 0$
modes.  This means that the imaginary part of the Fredholm integral
equation vanishes identically. Thus, there is just one equation and
one free parameter, which can in principle be adjusted to obtain
growing axisymmetric modes.  As the disk cannot distinguish between
modes with different growth rates, the self-consistent disks can admit
a one-dimensional continuum of growing axisymmetric modes.


\begin{figure}
	\begin{center}
	\centering\epsfig{file=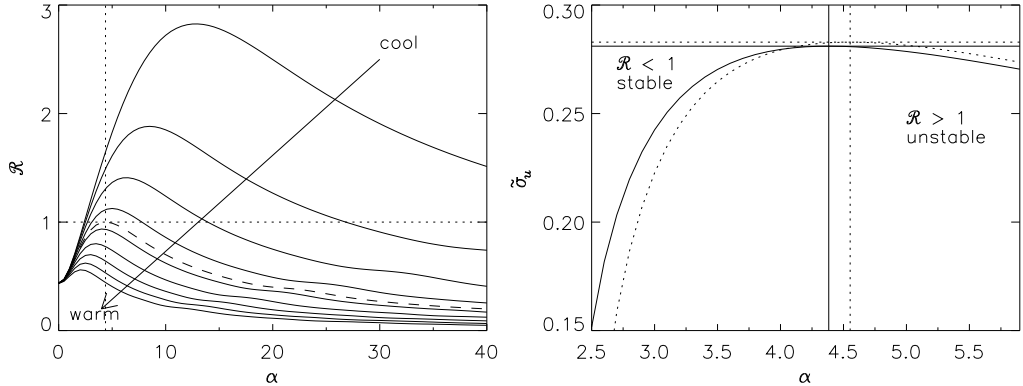,width=0.8\textwidth,height=0.3\textwidth}
	\caption [The global response function for $\beta = 0.25$. The
	marginal stability curve for the $\beta=0.25$ self-consistent
	disk] {The left panel shows the response function
	$\mathcal{R}(\alpha,\sigutil)$ for axisymmetric disturbances
	in the $\beta = 0.25$ disk. $\mathcal{R}$ is plotted as a
	function of wavenumber $\alpha$ for different values of the
	velocity dispersion $\sigutil$. The dashed line is the curve
	with $\sigutil=\sigutilmin$, that is, the critical velocity
	dispersion for stability according to local theory. The right
	panel shows the marginal stability curve $\mathcal{R}=1$ for
	the $\beta=0.25$ disk. The solid (dotted) lines indicate the
	result of the global (local) theory. [The temperature and
	wavenumber at which instability sets in are marked. These are
	$\sigutilmin=0.281$ for global stability, $\sigutilmin=0.283$
	for local stability; $\alphamu=4.38$ for global stability,
	$\alphamu=4.55$ for local stability].
	\label{fig:Selfconcurve_b0p25_globallocal}} \end{center}
\end{figure}

\begin{figure}
	\begin{center}
		\epsfig{file=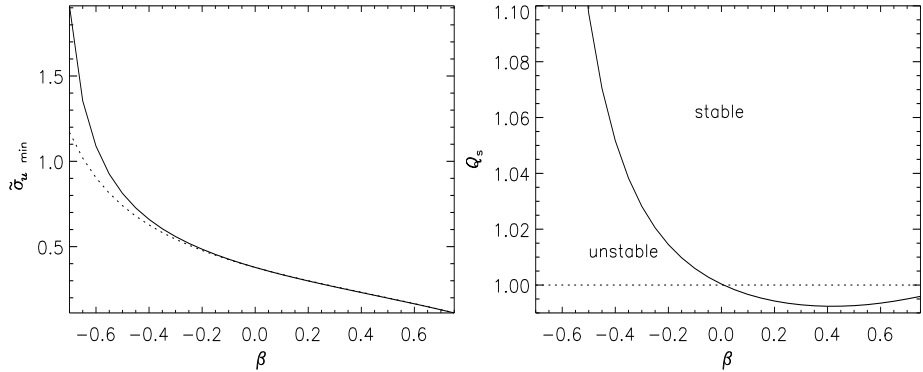,width=0.7\textwidth,height=0.3\textwidth}
		\caption [Minimum temperature for global axisymmetric
		stability plotted against rotation curve index $\beta$
		for a self-consistent disk] {Minimum temperature of
		the self-consistent power-law disk for global
		axisymmetric stability plotted against rotation curve
		index $\beta$. In the left-hand plot, this is
		presented as the minimum velocity dispersion
		$\sigutil$, and in the right-hand plot as the
		stability parameter $\Qsing$ necessary to stabilise
		the disk to axisymmetric perturbations. The dotted
		line shows the minimum temperature as given by local
		stability.
                \label{fig:m0marginalsing}} 
\end{center}
\end{figure}
\subsection{Neutral modes}\label{sec:neutral}
For the self-consistent disks, the neutral modes must have not just
$s=0$ but $\Omegap=0$ as well. As noted by Lynden-Bell \& Lemos
(1993), the neutral modes are pure logarithmic spirals. In this case,
the transfer function, which describes how the self-consistent disk
responds to forcing, can be streamlined from the cumbersome expression
given in Section 3.3 of Paper I to a simple delta-function. Let us
write $\Sm (\alpha, \alpha^\prime ) =
\mathcal{R} \delta ( \alpha - \alpha^\prime )$, where $\mathcal{R}$
defines the response function. We derive helpful expressions for the
response function in Appendix~\ref{sec:nmodes} and apply these
formulae in the following two sub-sections that study the axisymmetric
and bisymmetric neutral modes.

\subsubsection{Axisymmetric Neutral Modes}

\noindent
The only growing modes admitted by the self-consistent disk are
axisymmetric ones. These set in through the neutral modes (Lynden-Bell
\& Ostriker 1967; Kalnajs 1971).  The form of the response function
~\eqref{eq:mathcalRm0uv} in this instance is shown in the left panel
of Fig.~\ref{fig:Selfconcurve_b0p25_globallocal}.  Now, breathing
modes are long wavelength instabilities with no radial nodes. If
present, the response function crosses the self-consistent line
$\mathcal{R} =1$ as $\alpha \rightarrow 0$.
Fig.~\ref{fig:Selfconcurve_b0p25_globallocal} immediately makes clear
that the stellar power-law disks do not admit breathing modes -- in
contradistinction to the gas power-law disk~\cite{LKLB:1991}.
Breathing modes are analogous to the instability of stars with $\gamma
<4/3$, where $\gamma$ is the ratio of the heat capacity at constant
pressure to the heat capacity at constant volume (see Binney \&
Tremaine 1987, chap. 5). Breathing modes occur in gas disks because
gas molecules have internal degrees of freedom, which absorb energy
released in changes such as contraction. This energy is then not
available to the translational degrees of freedom which contribute to
the pressure. There is thus insufficient pressure to resist
gravitational collapse. Conversely, when a stellar disk contracts, all
the energy released contributes to increased random velocity, which
tends to counteract the effect of the increased gravity. Thus stellar
disks are much more stable than gas disks to breathing modes.

The marginal stability curve $\mathcal{R}=1$ is plotted in the plane
of dimensionless wavenumber $\alpha$ and velocity dispersion
$\sigutil$ in the right panel of
Fig.~\ref{fig:Selfconcurve_b0p25_globallocal}. This compares the
marginal stability curves obtained using local theory (dotted lines)
and global theory (solid lines). The agreement is very good,
especially close to the critical
temperature. Fig.~\ref{fig:m0marginalsing} shows the minimum
temperature needed for axisymmetric stability both in terms of the
velocity dispersion and the stability parameter $\Qsing$. It is clear
in this plot that the local and global results coincide most closely
at $\beta=0$, where the rotation curve is flat. This is probably
because Toomre's local analysis assumes that the stellar disk has a
Maxwellian velocity distribution. This is an excellent approximation
for the Toomre-Zang disks, less good for the other power-law disks.
Local theory becomes discrepant only when $\beta \lesssim
-0.25$. Toomre's (1964) local analysis assumes that the velocity
dispersion is small compared to the circular velocity, and that the
wavelength of any disturbance is small compared to the radius $R$ at
the point under consideration. In the present notation, these
conditions are $\sigutil \ll 1$ and $\alpha \gg 1$. They become harder
to justify when $\beta < -0.25$.  As $\beta$ becomes more negative,
the wavelength of the unstable perturbation becomes larger,
undermining local theory.  The local calculation then underestimates
the velocity dispersion needed for axisymmetric stability. Power-law
disks with rising rotation curves remain globally unstable even when
they are hot enough to be locally stable.
\begin{figure}
	\begin{center}
		\epsfig{file=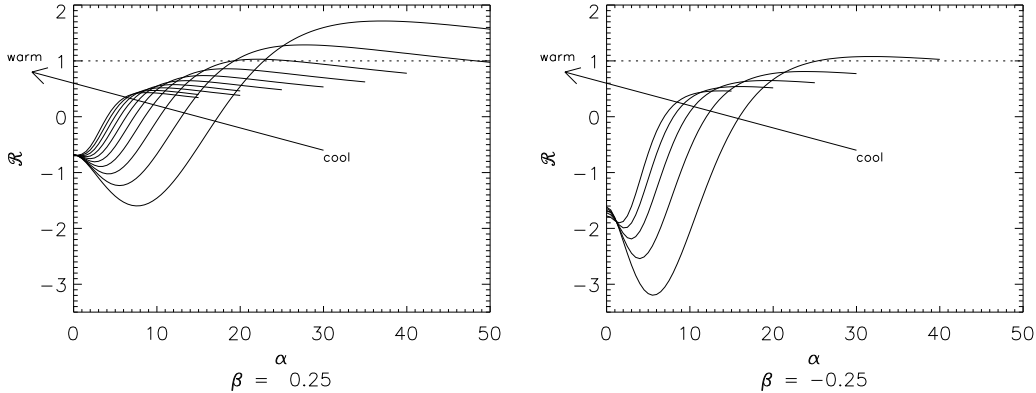,width=0.8\textwidth,height=0.3\textwidth}
		\caption[The neutral response function for
		self-consistent disks with $\beta = \pm 0.25$] {The
		response function for $m=2$ neutral modes in the
		self-consistent disks with $\beta = \pm 0.25$. [For
		$\beta=0.25$, curves are shown for ten values of
		$\Qsing$ in steps of 0.1 from 0.3 to 1.2. For
		$\beta=-0.25$, curves are shown for seven values of
		$\Qsing$ in steps of 0.1 from 0.3 to 0.9].
\label{fig:R_sing_m2}}
	\end{center}
\end{figure}

\subsubsection{Other Neutral Modes}

\begin{figure}
	\begin{center}
		\epsfig{file=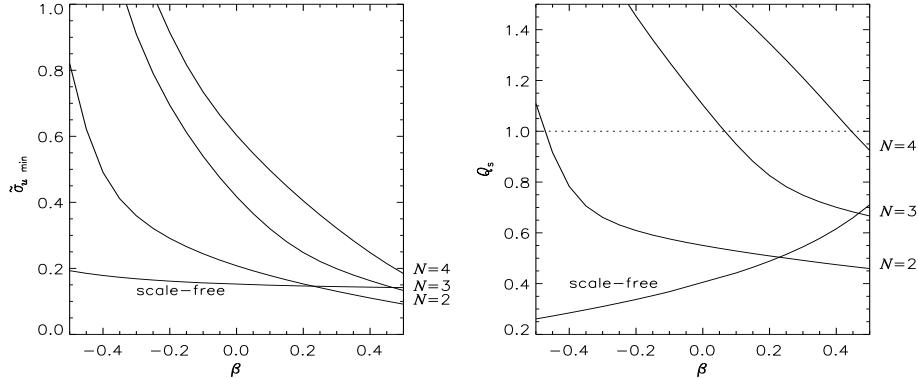,width=0.7\textwidth,height=0.3\textwidth}
		\caption[Minimum temperature for neutral modes plotted
		against $\beta$, for self-consistent and cut-out
		disks] {Velocity dispersion $\sigutil$ and stability
		parameter $\Qsing$ plotted against rotation curve
		index $\beta$ for an $m=2$ modes in self-consistent
		and cut-out disks. The curves labelled with the inner
		cut-out index $N$ show the $\sigutilmin$ and $\Qsing$
		at which the inner cut-out disks are marginally stable
		to bisymmetric disturbances. The curve labelled
		``scale-free'' shows the $\sigutilmin$ and $\Qsing$ at
		which the self-consistent disk just fails to admit any
		bisymmetric neutral modes.
\label{fig:Sing_m2_sigQ}}
	\end{center}
\end{figure}

Let us close with a brief examination of the bisymmetric neutral modes
admitted by the self-consistent disk.  The response function
$\mathcal{R}$ for these modes is given by
eq.~\eqref{eq:mathcalRnonaxuv}.  Fig.~\ref{fig:R_sing_m2} shows the
dependence of the response function on the wavenumber $\alpha$ for
different temperatures. It is difficult to obtain an accurate value
for $\mathcal{R}$ when both temperature and wavenumber are high. In
Fig.~\ref{fig:R_sing_m2}, the curve for each temperature has been
truncated when the inaccuracy becomes severe enough to be visible on
the graph.  When the temperature is low enough, the disk admits
neutral $m=2$ modes.  As the disk is heated, the
$\mathcal{R}$-$\alpha$ curve drops, until a critical temperature is
reached at which $\mathcal{R}=1$ only at some wavenumber. Above this
temperature, the response remains less than unity for all wavenumbers
and there are no neutral modes.  Fig.~\ref{fig:R_sing_m2} demonstrates
that disks with falling rotation curves admit neutral modes at higher
values of the stability criterion $\Qsing$ than disks with rising
curves. For $\beta=0.25$, the curve with $\Qsing=0.4$ intersects the
line $\mathcal{R}=1$, whereas for $\beta=-0.25$, this curve never
reaches $\mathcal{R}=1$.  The dependence of the critical temperature
for neutral modes on $\beta$ is shown in Fig.~\ref{fig:Sing_m2_sigQ}.
It has the opposite behaviour to marginally-stable modes in the
cut-out disks, which are shown on the same plot.

%% file: conc.tex
\section{Conclusions}
\noindent
The work in this paper concludes a linear normal mode analysis of a
family of self-gravitating, differentially rotating stellar dynamical
disks in which the circular velocity varies as a power of the
Galactocentric radius, i.e., $\vcirc \propto R^{-\beta/2}$. Our
analysis includes both self-consistent disks, in which the surface
density and the potential are related through Poisson's equation, and
cut-out disks, in which stars close to the centre (and sometimes also
at large radii) are carved out.  The analytic and numerical procedures
have exploited the self-similar force field of the models to develop a
fast code that finds the pattern speeds and growth rates of the normal
modes.  Some tables of the fastest growing modes are provided in
Appendix~\ref{sec:fastest}. A comparison of these linear stability
results with modes as detected in N-body simulations is a worthwhile
and interesting task.
%
%

\medskip
\noindent
(1) The inner cut-out is crucial in allowing growing non-axisymmetric
modes in the power-law disks. Physically, it is helpful to picture the
instabilities as standing wave patterns caused by the interference of
travelling waves in a resonant cavity.  For $m > 1$, the inner
Lindblad resonance absorbs incident waves and damps any disturbance.
Growing modes have pattern speeds which are large enough to push the
inner Lindblad resonance within the shield of the inner cut-out.
Waves cannot propagate easily within the core where there is little
active mass. The inner cut-out acts as a reflective barrier. So, it
has a destabilising effect if it shields waves from the absorbing
inner Lindblad resonance.  The inner cut-out is most effective as a
reflector of waves when it is abrupt. Disks with lower cut-out indices
allow more of the waves to seep through to the resonance and are thus
more securely stable. The $N=1$ disk has the mildest cut-out and is
the most stable of all.
 
\begin{figure}
	\begin{center}
		\epsfig{file=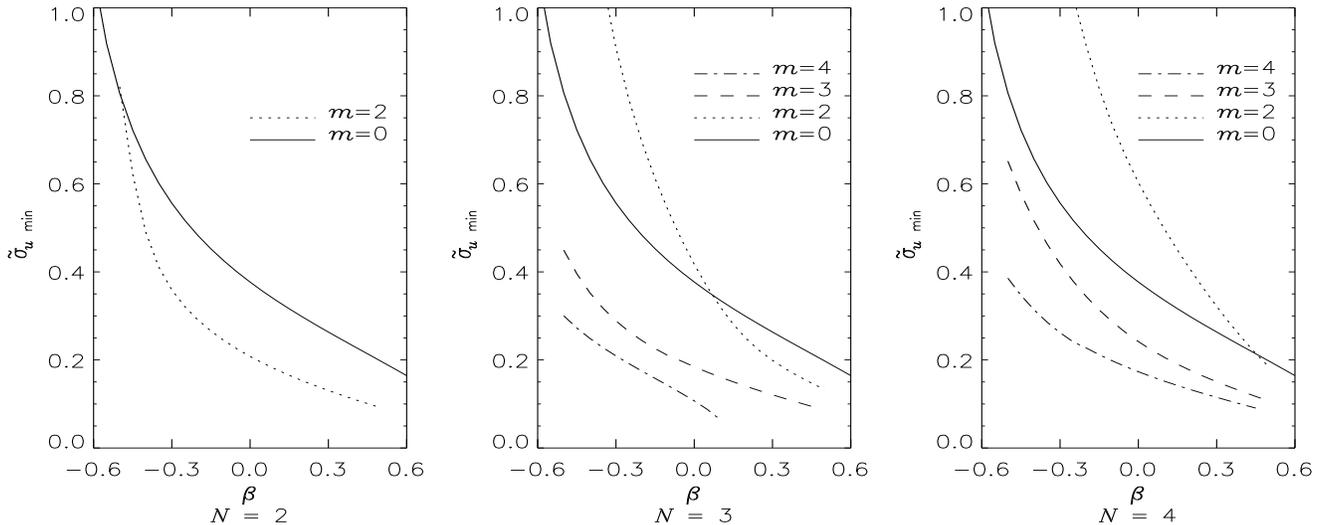,width=\textwidth,height=0.4\textwidth}
		\caption [Minimum velocity dispersion for global
		stability to modes with different azimuthal symmetry]
		{ Minimum temperature for global stability to modes
		with different azimuthal symmetry.
\label{fig:MarginalModes}
}
	\end{center}
\end{figure}

\medskip
\noindent
(2) The stability to different harmonics is summarised in
Fig.~\ref{fig:MarginalModes}, which shows the minimum velocity
dispersion needed for global stability. For an inner cut-out index $N
< 3$, the disks are more stable to bar-like or bisymmetric modes than
to axisymmetric disturbances, whereas for $N > 3$, the converse is
true.  Modes with $m=3$ and $m=4$ are always very hard to excite. The
disks must be extremely cold before they admit growing modes with more
than two arms.
The most serious instabilities to which all the cut-out disks are
susceptible are one-armed instabilities. At outset, let us concede
that our analysis permits the $m=1$ modes to shift the barycentre of
the galaxy (c.f. Zang 1976). This artefact arises from the rigid core
which has been carved out of the galaxy. A full treatment should allow
the core to move in response to the growing mode in such a way as to
cancel the shifting of the barycentre. Nonetheless, we do not believe
this to be a serious flaw -- first, because the barycentre moves only
slightly within the linear r\'egime and second, because N-body
simulations by~\longcite{Earn:1993}, using the code described
in~\longcite{ES:1995}, appear to corroborate the conclusion that
one-armed instabilities are the most serious.  The prevalence of the
one-armed modes can be easily understood at a simple level, as there
is no inner Lindblad resonance for $m=1$. This removes a powerful
stabilising influence.

What is the physical mechanism generating these one-armed modes?  They
are not akin to the edge
modes~\cite{Erickson:1974,Toomre:1981,Sell:1989}.  There is typically
only one edge mode characterised by a single growth rate and pattern
speed. By contrast, the cut-out disks exhibit an embarrassingly large
number of different one-armed unstable modes. Second, the one-armed
modes are not generated by refraction of short trailing waves into
long trailing waves (the `WASER') even though the $Q$ profile of the
cut-out disks does rise steeply in the inner regions in the manner
envisaged by Lin and co-workers~\cite{Mark:1976,LLM:1976}. A third
possibility, suggested to us by Sellwood (1997, private
communication), is that the cut-out may be essentially invisible to
long-wavelength disturbances, thus allowing trailing waves to leak
through the centre and emerge as leading waves with devastating
consequences. This does not seem to be the case, as the dominant
wavelengths of the unstable modes are not large compared to the
cut-out radius. Rather, the numerical evidence particularly of
Fig.~\ref{fig:m12_LeadingWaves} seems to us to suggest that the
trailing waves are reflected as leading waves at the inner cut-out. It
is this that completes the feedback for the swing-amplifier and causes
one-armed mayhem in the disk.

\medskip
\noindent 
(3) For all azimuthal wavenumbers, the unstable modes persist to
higher temperatures and grow more vigorously if the rotation curve of
the power-law disk is rising ($\beta <0$) rather than falling ($\beta
>0$). One way of understanding this result is to recall that, even in
local axisymmetric theory, the power-law disks with rising rotation
curves are more susceptible to Jeans instabilities and their ilk.  In
the absence of pressure, they remain vulnerable at longer
lengthscales.  They require larger velocity dispersions (relative to
the local circular speed) for stabilisation at short scales. For
example, on moving from $\beta = 0.5$ to $\beta = -0.5$, the typical
densities increase by a factor of roughly three (c.f., Table 1) and
this in itself indicates that instabilities of whatever sort are much
more worrisome in the models with rising rotation curves ($\beta <0$).
A number of authors~\cite{LH:1978,PapStwo:1991,Collett:1995} have
pointed out that gradients and especially turning points in the ratio
of vorticity to surface density (or potential vorticity $\zeta$) can
provoke instabilities.  In the power-law disks, the potential
vorticity behaves like
\be
\zeta = {\Omega \over \Sigma} \sim R^{-\beta/2}.
\end{equation}
The gradient of the potential vorticity actually vanishes in the
Toomre-Zang disk ($\beta =0$).  In the more nearly Keplerian disks
($\beta >0$), swing amplification is boosted by this effect. In the
more nearly uniformly-rotating disks ($\beta <0$), swing amplification
gets weakened. Somewhat surprisingly, the effect of gradient terms in
the potential vorticity actually runs contrary to the numerical
stability results. Any effect, though present, must be quite
weak. This has been demonstrated by Alar Toomre in unpublished work.

\medskip
\noindent
(4) The criterion suggested by Ostriker and Peebles (1973) for
stability against bar-like modes is untrustworthy. It gravely
overestimates the energy in the form of random motions necessary to
achieve stability against bisymmetric distortions. The power-law disks
are much more stable than the criterion suggests. Similar qualms have
already been reported by others~\cite{Aoki:1979,Toomre:1981}.

\medskip
\noindent
(5) There are no length-scales and time-scales in the scale-free
disks. If any mode is admitted at some pattern speed and growth rate,
then it must be present at all pattern speeds and growth rates. Here,
the analysis falls short of a complete proof, but our belief is that
such a two-dimensional continuum of non-axisymmetric modes does not
exist.  The weight of the evidence suggests that there is no continuum
and that the self-consistent disks admit no growing non-axisymmetric
modes whatsoever. There are two telling pieces of evidence in favour
of this conclusion. The first is the non-Hermitian nature of the
kernel in the Fredholm integral equation for non-axisymmetric modes,
which coerces us into satisfying (if we can) two simultaneous integral
equations with only one free parameter at our disposal. Second, the
behaviour of the mathematical eigenvalues of the cut-out disks in the
limit of vanishing pattern speed must mimic that of the
self-consistent disks and it does not suggest in the slightest the
existence of a continuum modes. At first sight, this conclusion is
astonishing -- as it applies even to the completely cold disk! But,
without reflecting boundaries such as provided by the cut-outs, there
is no resonant cavity and no possibility of unstable normal modes.
Such an outcome is already familiar in the theory of gaseous accretion
disks~\cite{PapS:1989,PapLin:1995}.

The case of the axisymmetric modes is somewhat different. Now, the
kernel is Hermitian and the eigenvalues must be real and
positive. There is only one integral equation to be solved with our
single disposable parameter. The temperature marking the onset of the
neutral modes is accurately given by local theory. On physical
grounds, these neutral modes are followed by growing modes as the
temperature of the disk is lowered further.  But the self-similar disk
cannot distinguish between modes with different growth rates. Below
the critical temperature the self-consistent disks must admit a
one-dimensional continuum of growing axisymmetric modes.

\medskip
\noindent
(6) The local theory of Toomre (1964) gives an excellent description
of the global axisymmetric stability of the cut-out and the
self-consistent power-law disks. The only possible axisymmetric modes
are Jeans modes. Breathing modes do not occur in the stellar power-law
disks, although they do occur in the gas disks~\cite{LKLB:1991}. This
is because gas molecules have internal degrees of freedom which can
absorb energy released in contraction. This energy is not available to
the translational degrees of freedom and does not contribute to the
gas pressure resisting collapse. By contrast, when a stellar disk
contracts, all the energy released is converted to random motions.

\medskip
\noindent
(7) Both the self-consistent and the cut-out disks admit neutral
modes. For the self-consistent disks, any neutral modes must have not
just $s=0$ but also $\Omegap =0$. The existence of neutral modes seems
to be related to the absence of $m$-lobed closed orbits in the frame
rotating with pattern speed $\Omegap$. The cleanest example of this
occurs for neutral triskele modes in the cut-out and self-consistent
disks. For such disks with $\beta \le -0.25$, there are neutral $m=3$,
$\Omegap=0$ modes independent of temperature. These are the very disks
for which three-lobed orbits closed in the inertial frame do not
occur. This is straightforward to understand. The existence of any
three-lobed closed orbits in the $\Omegap =0$ frame would cause any
neutral pattern to evolve.

\bigskip
\noindent
NWE and JCAR are particularly indebted to David Earn and Jerry
Sellwood for the loan of computer code that simulates the evolution of
the cut-out power-law disks using the smooth-field-particle code
described in David Earn's Ph. D. thesis. In particular, David Earn
kindly re-ran some of the models in his thesis so that the dominant
unstable modes could be checked against our linear stability code. We
wish to acknowledge valuable conversations with James Binney, Jim
Collett, David Earn, Jeremy Goodman, Donald Lynden-Bell, Jerry
Sellwood, Alar Toomre and Scott Tremaine -- all of whose insights have
benefitted us enormously. Both Jerry Sellwood and Alar Toomre offerred
extensive and useful comments on earlier versions of the manuscript.
Finally, NWE and JCAR thank the Royal Society and the Particle Physics
and Astronomy Research Council respectively for financial support.

%% file: appa.tex
This appendix briefly discusses the global stability of the cut-out
power-law disks to axisymmetric modes.  In the axisymmetric case, the
marginal modes demarking the boundary between stability and
instability are those of zero growth rate and pattern
speed~\cite{LBOst:1967,Kalnajs:1971}. In fact, the limit of vanishing
growth rate gives rise to some numerical delicacies in the calculation
of the kernel of the Fredholm integral equation.  The expression for
the cut-out angular momentum function $\Flm$ is derived in Appendix C
of Paper I. For $m=0$, it becomes
\begin{equation}
	\lim_{\stil\rightarrow 0^+} F_{l0}(\eta \not= 0) 
	=
	- \twomp \modterm
\frac{i}{1-e^{2\pi\hat{\eta}}}
\left\{
	\lim_{\stil\rightarrow 0^+} e^{-i\hat{\eta} \ln \frac{l\kaptil}{i\stil}}
+
	\frac{( 1 - e^{\pi\hat{\eta}} ) }{N( 1 - e^{-\pi\hat{\eta}/N} ) }
\right\}.
	\label{eq:W008}
\end{equation}
Taking the principal value of the logarithm, we find that
\be
	\ln {{l\kaptil} \over {i\stil} } = \ln \left| {{l\kaptil}
	\over \stil} \right| + i\pi \left(1 + \text{sign} (l) \times
	\half\right),
\end{equation}
and hence the limit in~\eqref{eq:W008} is
\be
	\lim_{\stil\rightarrow 0^+} e^{-i\hat{\eta} \ln
	 \frac{l\kaptil}{i\stil}} = e^{\pi\hat{\eta}
	 (1+\text{sign}(l)\times \half)} \lim_{\stil\rightarrow 0^+}
	 e^{-i\hat{\eta} \ln |l\kaptil/\stil | }, \label{eq:W009}
\end{equation}
\begin{figure}
	\begin{center}
		\epsfig{file=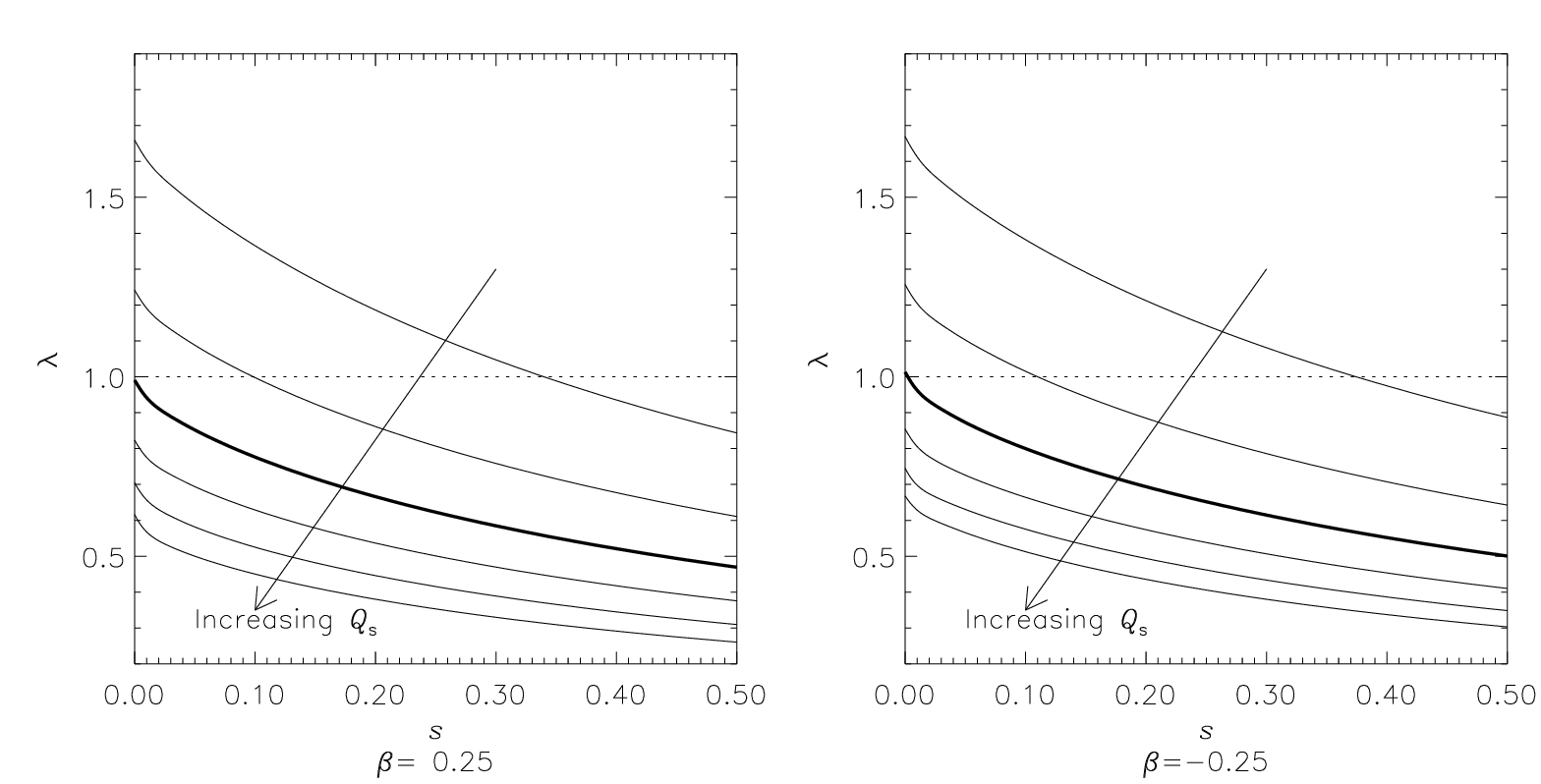,width=0.775\textwidth,height=0.325\textwidth}
		\caption [Dependence of the largest mathematical
		eigenvalue on growth rate and temperature] {Dependence
		of the largest mathematical eigenvalue on growth rate
		$s$ and temperature $\Qsing$ for axisymmetric
		disturbances in inner cut-out disks with
		$N=2$. [Curves are drawn for six different values of
		$\Qsing$: $\Qsing=0.6$, 0.8, 1.0, 1.2, 1.4, 1.6. The
		$\Qsing=1$ curve is drawn with a bolder line. For the
		$\beta=0.25$ disk, $\Qsing=1$ corresponds to a
		velocity dispersion of $\sigutil=0.283$, and for
		$\beta=-0.25$, to $\sigutil=0.509$].
		\label{fig:m0eigen_sQ}} \end{center}
\end{figure}
For very negative values of $\eta$, the $e^{\pi\eta}$ term
in~\eqref{eq:W009} ensures that the limit is zero. But for positive
$\eta$, the expression for $\Flm$ fails to converge to any definite
limit. This might be expected to present problems for the numerical
implementation. Fortunately, it turns out that the mathematical
eigenvalue converges even though $\Flm$ does not (see Read 1997 for
ample numerical evidence). This is because -- especially when the
growth rate is small -- the eigenvalue is dominated by the matrix
elements along the diagonal $\eta=0$, where the troublesome term is
simply unity.  In eq.~\eqref{eq:W008}, we can write $N ( 1- e^{-\pi
\hat{\eta} / N} ) \approx \pi \hat{\eta}$ close to $\eta=0$. The
dependence on the inner cut-out function $N$ is cancelled out in the
matrix elements close to the diagonal, where the transfer function is
largest. We thus expect the stability of the inner cut-out disk to be
virtually independent of $N$ -- as already argued on physical grounds.

\begin{figure}
	\begin{center}
		\epsfig{file=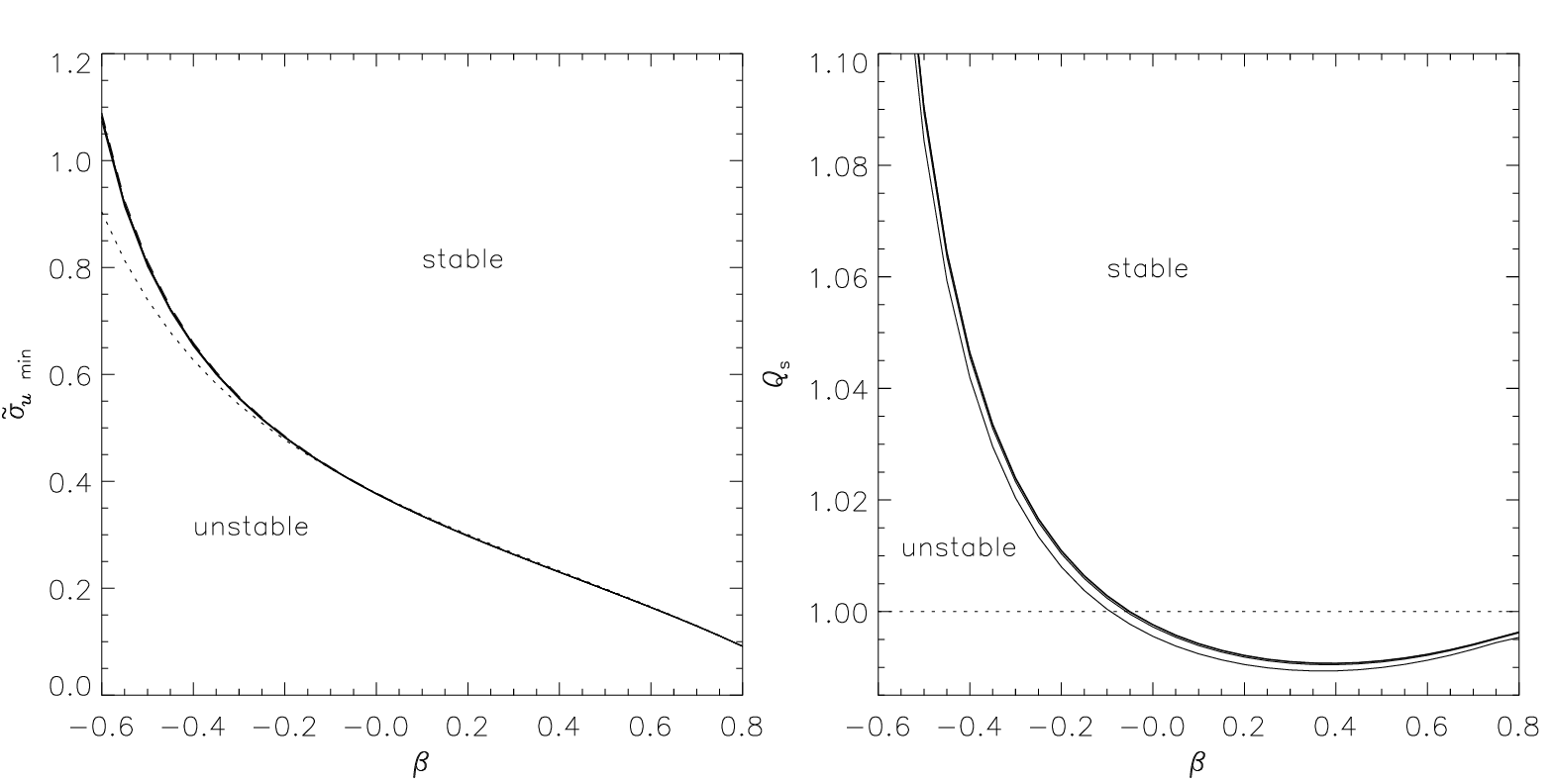,width=0.75\textwidth,height=0.3\textwidth}
		\caption[Minimum temperature for global axisymmetric
		stability plotted against $\beta$ for cut-out disks]
		{Minimum temperature for global axisymmetric stability
		plotted against rotation curve index $\beta$ for disks
		with various cut-out functions. The left panel shows
		the minimum velocity dispersion $\sigutil$; the right
		panel shows the same data in terms of the stability
		parameter $\Qsing$. The dotted lines show the local
		theory; other lines indicate results of global theory
		for an inner cut-out disks. [Results for $N=1$, 2, 3,
		4 are plotted in both graphs, but coincide so closely
		as to be virtually indistinguishable.]
		\label{fig:betagamma_m0_cutout}} \end{center}
\end{figure}

Fig.~\ref{fig:m0eigen_sQ} shows how the mathematical eigenvalue
depends on growth rate and temperature for two sample cut-out disks.
For sufficiently small $\Qsing$, a unit eigenvalue can always be found
for large enough growth rate $s$. When the disk is cool, it is
susceptible to fiercely growing modes. As the disk is heated, the
growth rate of the unstable mode decreases. The temperature at which
the disk is marginally stable is given by the $\Qsing$ for which the
$s=0^+$ eigenvalue is unity.  Inspection of Fig.~\ref{fig:m0eigen_sQ}
shows that this value of $\Qsing$ is close to unity. This already
suggests that the global axisymmetric stability of the cut-out disks
is close to the local axisymmetric
stability. Fig.~\ref{fig:betagamma_m0_cutout} shows the velocity
dispersion $\sigutilmin$ and the stability parameter $\Qsing$ of the
marginally stable disks as a function of rotation curve index
$\beta$. Four different values of $N$ are plotted, but the curves are
scarcely distinguishable. For comparison, the results of local theory
for the self-consistent disk are shown in a dotted line.
Local theory is an excellent guide to the global stability of the
inner cut-out disks. This is in accordance with Toomre and Zang's
findings -- for the disk with a flat rotation curve, local and global
results agree ``to within a few tenths of one per
cent''~\cite{Zang:1976}. The value of the inner cut-out index has
little effect on the stability. An additional outer cut-out function
tends to make the disk slightly more stable, in that a lower velocity
dispersion is needed for axisymmetric stability.

%% file: appb.tex
In this appendix, we demonstrate that if any modes with $\omega \ne 0$
exist in the self-consistent disk, then they must form a
two-dimensional continuum. This proof is a straightforward extension
of that due to Kalnajs in 1974 for the disk with a flat rotation curve
(given in Zang's (1976) thesis).

For the scale-free disk, the angular momentum function is defined as
\begin{equation}
 	 \Flm (\eta)
 	= 
	\left\{ ( \lkapmom ) \modterm - \gamma m \right\}
	{ 1 \over {2\pi} } \int_0^\infty
	{ {e^{-i \eta {2\over {2-\beta}}  \ln \Ltil }}
	 \over { \lkapmom -\omtil \Ltil^{\twopm}} }
	 { {d\Ltil}  \over \Ltil}.
	\label{eq:FlmSingDisk}
\end{equation}
where the notation of Paper I is used. As shown in Appendix C of
Paper I, the angular momentum function consists of a regular function
and a Dirac delta-function. Here, let us write:
\be
	\Flm (\eta )
	=
	\Flm^{(1)} (\eta)
	+ 
	 \Flm^{(2)}  \delta(\eta )
	\label{eq:005}
\end{equation}
where
\begin{equation}
	\Flm^{(1)} (\eta)
	 =
	 - \twomp 
	\left\{ \modterm - \frac{\gamma m}{\lkapmom} \right\}
	\frac
	 {i e^{-i \hat{\eta} \ln { \lkapmom \over \omtil}} }
	 { 1- e^{2\pi\hat{\eta}} } ,
	\label{eq:Flm1sing}
\end{equation}
\begin{equation}
	\Flm^{(2)} 
	 = 	
	\frac{2-\beta}{4}
	\left\{  \modterm - \frac{\gamma m}{\lkapmom}  \right\}	,
	\label{eq:Flm2sing}
\end{equation}
and $\hat{\eta} = 2\eta /(2+\beta)$.  (For $m \ne 0$, these
expressions hold for all $l$. For $m=0$, there is an ambiguity at the
$l=0$ harmonic. The correct choices are $F_{00}^{(1)} = F_{00}^{(2)} =
0$).  When $\omega=0$, the angular momentum function is simply a
delta-function in $\eta$:
\begin{equation}
 	 \Flm (\eta)
	 = 	\left\{ \modterm - \frac{\gamma m}{\lkapmom} \right\}
	 \int_0^\infty
	 \frac{e^{-i \eta {2\over {2-\beta}}  \ln \Ltil }}{2\pi}
	 { {d\Ltil}  \over \Ltil}
	=
	\frac{2-\beta}{2}
	\left\{ \modterm - \frac{\gamma m}{\lkapmom} \right\} \delta(\eta).
	\label{eq:FlmSingOmEqZero}
\end{equation}
When $\omega=0$, we do not pick up the ``off-diagonal'' terms
describing the response at wavelengths different from that of the
original perturbation.  Let us note that $\Flm^{(1)} (\eta)$ has no
unique limiting value as $\omega \rightarrow 0$. It does not tend to
zero in this limit and the case $\omega=0$ cannot be recovered by
taking the limit $\omega \rightarrow 0$.  Physically, this is a
consequence of the symmetry of the disk.  A pure logarithmic spiral is
self-similar and a neutral logarithmic spiral introduces no length- or
time-scales through its growth rate and pattern speed. Thus the
response must be purely self-similar, i.e., another logarithmic
spiral.  Once the logarithmic spiral is allowed to grow or rotate --
however slowly -- a length- and time-scale has been introduced. The
response is no longer self-similar, but involves contributions at many
wavelengths. 

Analogously to~\eqref{eq:005}, let us divide the transfer function
into a regular part and a Dirac delta-function:
\begin{equation}
	\Sm ( \alpha, \alpha^\prime )
	=
	\mathcal{S}_m^{(1)} ( \alpha, \alpha^\prime )
	+
	\delta(\alpha - \alpha^\prime ) 
	\mathcal{S}_m^{(2)}  ( \alpha, \alpha^\prime)
	\label{eq:018}.
\end{equation}
$\mathcal{S}_m^{(1)} ( \alpha, \alpha^\prime )$ and
$\mathcal{S}_m^{(2)} ( \alpha, \alpha^\prime )$ may be obtained from
eq. (78) of Paper I, using $\Flm^{(1)}$ and $\Flm^{(2)}$ respectively
in place of $\Flm$. We see that $\Flm^{(1)} (\eta)$ admits the
factorisation
\begin{equation}
	\Flm^{(1)}(\eta) = e^{i\hat{\eta}\ln\omtil} \hat{F}_{lm}^{(1)} (\eta),
	\label{eq:wyn1}.
\end{equation}
where $\hat{F}_{lm}^{(1)} (\eta)$ is independent of the growth
rate and pattern speed. The transfer function depends on the
growth rate only through the angular momentum function. So,
$\mathcal{S}_m^{(2)} ( \alpha, \alpha^\prime )$ is independent of
$\omtil$, while the dependence of $\mathcal{S}_m^{(1)}( \alpha,
\alpha^\prime )$ on $\omtil$ can be factorised out as
\be
	 \mathcal{S}_m^{(1)}( \alpha, \alpha^\prime ) 
	= e^{i (\hat{\alpha} - \hat{\alpha}^\prime) \ln \omtil } 
	\hat{\mathcal{S}}_m^{(1)} ( \alpha, \alpha^\prime ),
\end{equation}
where $\hat{\alpha}$ is defined by $\hat{\alpha}= 2\alpha /(2+\beta)$
and $\hat{\mathcal{S}}_m^{(1)} (\alpha, \alpha^\prime )$ is
independent of $\omtil$.  Let us define reduced transforms via:
\be
	\hat{A} (\alpha) = e^{-i \hat{\alpha} \ln \omtil } A (\alpha).
	\label{eq:Ahat}
\end{equation}
Then, the Fredholm integral equation becomes
\be
	 \lambda \hat{A} ( \alpha ) 
	=
	\hat{A} ( \alpha ) \mathcal{S}_m^{(2)} (\alpha, \alpha )
	+
	\int_{-\infty}^{+\infty} d\alpha^\prime
	\hat{A} ( \alpha^\prime ) 
	\hat{\mathcal{S}}_m^{(1)} ( \alpha, \alpha^\prime ) 
	,
	\label{eq:inteqnsing}
\end{equation}
where none of the quantities depend on $\omtil$. This means that if a
non-trivial solution $\hat{A} (\alpha )$ exists at all, then
self-consistent modes $A(\alpha)$ exist with any non-vanishing growth
rate and pattern speed.

%% file: appc.tex
In this appendix, expressions are derived for the response of the
self-consistent disk to elementary forcing by a pure logarithmic
spiral. For the neutral modes, the angular momentum function is simply
a delta-function~\eqref{eq:FlmSingOmEqZero}.  Thus, the transfer
function is also proportional to a delta-function. Let us write $\Sm (
\alpha, \alpha^\prime ) =
\mathcal{R} \delta ( \alpha - \alpha^\prime)$, where 
$\mathcal{R}$ is the response function:
\begin{equation}
\begin{split}
	\mathcal{R} & = \left( 1 - \frac{\beta}{2} \right) 2 \pi G
K(\alpha,m) \tilde{C} R_0^{\gamma+1} v_\beta^{{2 \over \beta}
(1+\gamma) } \int d\Util \, \auxint_0(\Util) \Util
\\ & \times 
	\modenergy ^{{1 \over \beta} + {\gamma \over \beta} - {\gamma \over 2}}
	\sum_{l=-\infty}^{+\infty} |\Qlm ( \alpha )|^2
	\left\{ \modterm - \frac{\gamma m}{\lkapmom} \right\} 
	\label{eq:mathcalRnonax}.
\end{split}
\end{equation}
The response density transform $\Ares (\alpha)$ at any wavenumber
$\alpha$ is simply proportional to the imposed density transform
$\Aimp (\alpha)$. Physically, the ratio of the response density to the
imposed perturbation is given by the response function
$\mathcal{R}$. If $\mathcal{R}>1$, the perturbation grows; if
$\mathcal{R}<1$, the perturbation dies away. Self-consistent neutral
modes require $\mathcal{R}=1$. We note that $\mathcal{R}$ is real and
independent of the radial position in the disk, so that the disk is
either stable or unstable everywhere.  Recalling the symmetry
properties of the Kalnajs function and of the Fourier coefficients, it
is evident that the response function is independent of the sign of
the wavenumber. This is an expression of the anti-spiral
theorem~\cite{LBOst:1967}. Every spiral of wavenumber $\alpha$ comes
with an ``anti-spiral'' of wavenumber $-\alpha$. Hence any neutral
modes come in pairs, one leading and one trailing.

There is an alternative form of the response
function~\eqref{eq:mathcalRnonax}, which merely involves integration
over one period of the unperturbed orbit. This is especially useful
for the important case of the $m=0$ neutral modes, as it resolves the
ambiguity at the $l=0$ harmonic.  In this derivation, the response to
a pure logarithmic spiral is itself assumed to be a pure logarithmic
spiral. According to Zang (1976), this calculation was carried out for
the axisymmetric neutral modes of the disk with a completely flat
rotation curve by Toomre in the 1960s. Putting together eqs. (51),
(58) and (68) of Paper I, we obtain for the change in the distribution
function
\begin{equation}
	\fimp (t) 
	= 
	\tilde{C}L_z^{\gamma-1}	 
	| E | ^{\betagammabig } 
	\psimp^{\alpha m} (t) 
\left\{
	\left| \betagammabig \right|  
	\frac{L_z}{|E|}
	+ i
\left(
	\left| \betagammabig \right|  
	\frac{L_z}{|E|}
	\omega 
	-  
	m \gamma   
\right)
	\int_{-\infty}^{t} \frac{\psimp^{\alpha m} \left( t^\prime \right)}
	{\psimp^{\alpha m} ( t )} dt^{\prime}.
\right\}
\end{equation}
Substituting the potential of a single logarithmic spiral, we obtain
\begin{equation}
\begin{split}
&	\fimp (t) 
	= 
	2 \pi G \tilde{C} \Sigmap K(\alpha,m) R_0
	L_z^{\gamma-1}	 
	\left| E \right| ^{\betagammasmall } 
	e^{ i \left( m \theta - \omega t \right) }
	\left( {R \over {R_0}} \right) ^{ i  \alpha - \half }
\left\{
	\left|\betagammabig \right|  
	\frac{L_z}{|E|}
\right.
\\&
\left.	+ 
i\left(
	\left| \betagammabig \right|  
	\frac{L_z}{|E|}
	 \omega 
	-  
	m \gamma   
\right)
	\int_{-\infty}^{t} 	
	e^{ i \left( m (\theta^\prime-\theta) - \omega (t^\prime-t) \right) }
	\left( {R^\prime \over {R}} \right) ^{ i  \alpha - \half }
	 dt^\prime 
\right\}.
\end{split}
\end{equation}
The integral in this equation can be split up into an infinite
sequence of integrals over the radial period $T$. Summing the
geometric progression, we have
\begin{equation}
\begin{split}
&	\fimp (t) 
	=
 	2 \pi G \tilde{C} \Sigmap K(\alpha,m) R_0
	L_z^{\gamma-1}	 
	| E | ^{ \betagammasmall} 
	e^{ i \left( m \theta - \omega t \right) }
	\left( {R \over {R_0}} \right) ^{ i  \alpha - \half }
\left\{
	\left| \betagammabig \right|  
	\frac{L_z}{|E|} 
\right.
\\&
\left.+
i\left(
	\left|\betagammabig \right|  
	\frac{L_z}{|E|}
	 \omega 
	-  
	m \gamma   
\right)
	\frac{1}{1 - e^{-iT(m\Omega-\omega)} }
	\int_{t-T}^{t} 	
	e^{ i \left( m (\theta^\prime-\theta) - \omega (t^\prime-t) \right) }
	\left( {R^\prime \over {R}} \right) ^{ i  \alpha - \half }
	 dt^\prime
\right\}.
	\label{eq:054}
\end{split}
\end{equation}
For $m \ne 0$, we simply set $\omega=0$ to obtain
\begin{equation}
\begin{split}
	\fimp (t) 
&	= 	2 \pi G \tilde{C} \Sigmap K(\alpha,m) R_0
	L_z^{\gamma-1}	 
	| E | ^{\betagammasmall } 
	e^{ i m \theta  }
	\left( {R \over {R_0}} \right) ^{ i  \alpha - \half }
\\&\times
\left\{
	\left| \betagammabig \right|  
	\frac{L_z}{|E|}
	-  
	\frac{im \gamma  }{1 - e^{-im\Omega T} }
	\int_{t-T}^{t} 	
	e^{ i  m (\theta^\prime-\theta)}
	\left( {R^\prime \over {R}} \right) ^{ i  \alpha - \half }
	 dt^\prime
\right\}.
\end{split}
\end{equation}
This expression can then be integrated over velocity space to obtain
the response density $\Sigres$. Recalling the definition of the response 
function $\mathcal{R}$, we finally obtain
\begin{equation}
\begin{split}
	\mathcal{R} 
& =	
	2 \pi G \tilde{C}  K(\alpha,m) R
	\iint du\, dv\, 
	L_z^{\gamma-1}	 
	| E | ^{ \betagammasmall } 
\\& \times
\left\{
	\left| \betagammabig \right|  
	\frac{L_z}{|E|}
	-  
	\frac{im \gamma  }{1 - e^{-im\Omega T} }
	\int_{t-T}^{t} 	
	e^{ i  m (\theta^\prime-\theta)}
	\left( {R^\prime \over {R}} \right) ^{ i  \alpha - \half }
	 dt^\prime
\right\}.
	\label{eq:mathcalRnonaxuv}
\end{split}
\end{equation}
This expression can be shown to be equivalent to that obtained 
directly from the Fredholm integral equation (see Read 1997).

When the imposed perturbation is axisymmetric, eq.~\eqref{eq:054}
is singular, viz.,
\begin{equation}
\begin{split}
	\fimp (t) 
&	= 	2 \pi G  K(\alpha,0)\tilde{C} \Sigmap R_0
	e^{ s t }
	\left( {R \over {R_0}} \right) ^{ i  \alpha - \half }
	L_z^{\gamma-1}	 
	| E | ^{\betagammasmall } 
\\& \times
\left\{
	\left| \betagammabig \right|  
	\frac{L_z}{|E|} 
\right.
\left.-
	\left| \betagammabig \right|
	\frac{L_z}{|E|}
	\frac{s}{1 - e^{-sT} }
	\int_{t-T}^{t} 	
	e^{ s (t^\prime-t)  }
	\left( {R^\prime \over {R}} \right) ^{ i  \alpha - \half }
	 dt^\prime
\right\}
\end{split}
\end{equation}
Using l'H\^{o}pital's rule to take the limit as $s \rightarrow 0$,
we find
\begin{equation}
	\fimp (t) 
	= 
	2 \pi G K(\alpha,0) \tilde{C} \Sigmap R_0
	\left| \betagammabig \right|
	\left( {R \over {R_0}} \right) ^{ i  \alpha - \half }
	L_z^{\gamma}	 
	| E | ^{\betagammasmall -1} 
\left\{ 1
	-
	\frac{1}{ T  }
	\int_{t-T}^{t} 	
	\left( {R^\prime \over {R}} \right) ^{ i  \alpha - \half }
	 dt^\prime
\right\}.
\end{equation}
This can be integrated over velocity to obtain the axisymmetric response
function
\begin{equation}
	\mathcal{R}
	=
 	2 \pi G  K(\alpha,0)  \tilde{C}
	\left| \betagammabig \right|
	R
	\iint du\,dv\,
	L_z^{\gamma}	 
	| E | ^{\betagammasmall -1} 
\left\{ 1
	-
	\frac{1}{ T  }
	\int_{t-T}^{t} 	
	\left( {R^\prime \over {R}} \right) ^{ i  \alpha - \half }
	 dt^\prime
\right\}.
	\label{eq:mathcalRm0uv}
\end{equation}
In practice, these forms of the response function for neutral modes
are more useful than~\eqref{eq:mathcalRnonax} because they involve
a quadrature performed over just one radial oscillation of the orbit.

%% file: appd.tex
The following sets of tables compare the stability of inner cut-out
disks to perturbations of different azimuthal symmetry $m$.  For each
symmetry $m$ and inner cut-out index $N$, the pattern speed and growth
rate of the fastest-growing mode are recorded.  The results for
$\beta=0.25$, $\beta=0.00$ and $\beta=-0.25$ are shown in different
tables. The numerical accuracy parameters are: $n = 301$,
$\Delta\alpha = 0.1$, $l_{\min} = -30$, $l_{\max} = +40$, $n_{GL} =
9$, $f_\sigma = 0.8$, $a_{\text{acc}} = 20$, $b_{\text{acc}} = 2$,
$\epsilon_\lambda=10^{-6}$. The data in the tables are also presented
in graphical form in the Figs~D1 and D2.  The pattern speed of the
fastest-growing mode depends almost linearly on $N$.  The fastest
growth rate increases quite steeply with $N$, reflecting the increased
instability of the more sharply cut-out disks.

  \begin{center}
  \begin{tabular}
  {|p{0.4cm}|p{1.4cm}p{1.6cm}|p{1.4cm}p{1.6cm}|p{1.4cm}p{1.6cm}|}
  \hline 
\multicolumn{7}{|c|}{$\Qsing=1.0$, $\beta=0.25$: $\sigutil = 0.283$, $\gamma = 11.0$}
\\ \hline
  & \multicolumn{2}{|c|} { $m=1$ }
  & \multicolumn{2}{|c|} { $m=2$ }
  & \multicolumn{2}{|c|} { $m=3$ }
  \\
  $N$  & $\Omegap$ & $s$ & $\Omegap$ & $s$ & $\Omegap$ & $s$\\ \hline
  1
 	& 0.097362	&	0.014041
 	& 	-	&	-	
 	& 	-	&	-	
\\ 
  2
 	& 0.157969	&	0.050158
 	& 	-	&	-	
 	& 	-	&	-	
\\ 
  3
 	& 0.192207	&	0.078881
 	& 	-	&	-
 	& 	-	&	-
\\ 
  4
 	& 0.213645	&	0.099969
 	& 0.460196	&	0.060416
 	& -		&	-
\\ 
  5
 	& 0.228252	&	0.115400	
	& 0.466727	&	0.128279
 	& -		&	-
\\ 
  6
 	& 0.238798	&	0.126749
 	& 0.472435	&	0.170714
 	& 	-	&	-
\\ 
  7
 	& 0.246718	&	0.135160
 	& 0.477593	&	0.199573
 	& 0.619188	&	0.016944
\\ 
  8
 	& 0.252824	&	0.141454
 	& 0.482190 	&	0.220295
 	& 0.623895	&	0.076268
\\ 
\hline
\end{tabular}
\end{center}

  \begin{center}
  \begin{tabular}
  {|p{0.4cm}|p{1.4cm}p{1.6cm}|p{1.4cm}p{1.6cm}|p{1.4cm}p{1.6cm}|}
  \hline 
\multicolumn{7}{|c|}{$\Qsing=1.0$, $\beta=0.00$: $\sigutil = 0.378$, $\gamma = 6.00$}
\\ \hline
  & \multicolumn{2}{|c|} { $m=1$ }
  & \multicolumn{2}{|c|} { $m=2$ }
  & \multicolumn{2}{|c|} { $m=3$ }
  \\
  $N$  & $\Omegap$ & $s$ & $\Omegap$ & $s$ & $\Omegap$ & $s$ \\ \hline
  1
 	& 0.085847	&	0.024053
 	& 	-	&	-	
 	& 	-	&	-	
\\ 
  2
 	& 0.140888	&	0.065923
 	& 	-	&	-	
 	& 	-	&	-	
\\ 
  3
 	& 0.170997	&	0.096969
	& 0.433518   	&	0.021843
 	& 	-	&	-	
\\ 
  4
	& 0.189628	&	0.119035
	& 0.439460	&	0.127192
 	& 	-	&	-
\\ 
  5
	& 0.202258	&	0.134777
 	& 0.445317	&	0.185191
 	& -	&	-
\\ 
  6
	& 0.211327	&	0.146108
	& 0.451135	&	0.222089
 	& 0.600215	&	0.006528
\\ 
  7
 	&  0.218103	&	0.154365	
 	&  0.456487	&	0.247575
 	&  0.605870 	&	0.081236
\\ 
  8
 	& 0.223301	&	0.160461
 	& 0.461228	&	0.266062
 	& 0.610127	&	0.134753
\\ 
\hline
\end{tabular}
\end{center}

  \begin{center}
  \begin{tabular}
  {|p{0.4cm}|p{1.4cm}p{1.6cm}|p{1.4cm}p{1.6cm}|p{1.4cm}p{1.6cm}|}
  \hline 
\multicolumn{7}{|c|}{$\Qsing=1.0$, $\beta=-0.25$: $\sigutil = 0.509$, $\gamma = 3.36$}
\\ \hline
  & \multicolumn{2}{|c|} { $m=1$ }
  & \multicolumn{2}{|c|} { $m=2$ }
  & \multicolumn{2}{|c|} { $m=3$ }
  \\
  $N$  & $\Omegap$ & $s$ & $\Omegap$ & $s$ & $\Omegap$ & $s$ \\ \hline
  1
 	& 0.080733	&	0.034289
 	& 	-	&	-	
 	& 	-	&	-	
\\ 
  2
 	& 0.132275	&	0.081594
 	& 	-	&	-	
 	& 	-	&	-	
\\ 
  3
 	& 0.160210	&	0.115238	
 	& 0.423391	& 	0.102223
 	& 	-	&	-	
\\ 
  4
	& 0.177418	&	0.138674
	& 0.427591	&	0.191581
 	& 	-	&		-
\\ 
  5
 	& 0.189107	&	0.155207
 	& 0.433935	& 	0.242558
 	& 	-	&		-
\\ 
  6
	&  0.197443	&	0.166961
 	&  0.440325	&	0.276073
 	&  0.599256	&	0.141945
\\ 
  7
 	&  0.203663	&	0.175475
	&  0.446097	&	0.299705
	&  0.599256	&	0.141945
\\ 
  8
 	&  0.208430	&	0.181740
	&  0.451131	&	0.317072
	&  0.603496	& 	0.190117
\\
\hline
\end{tabular}
\end{center}

\begin{figure}
	\begin{center}
		\epsfig{file=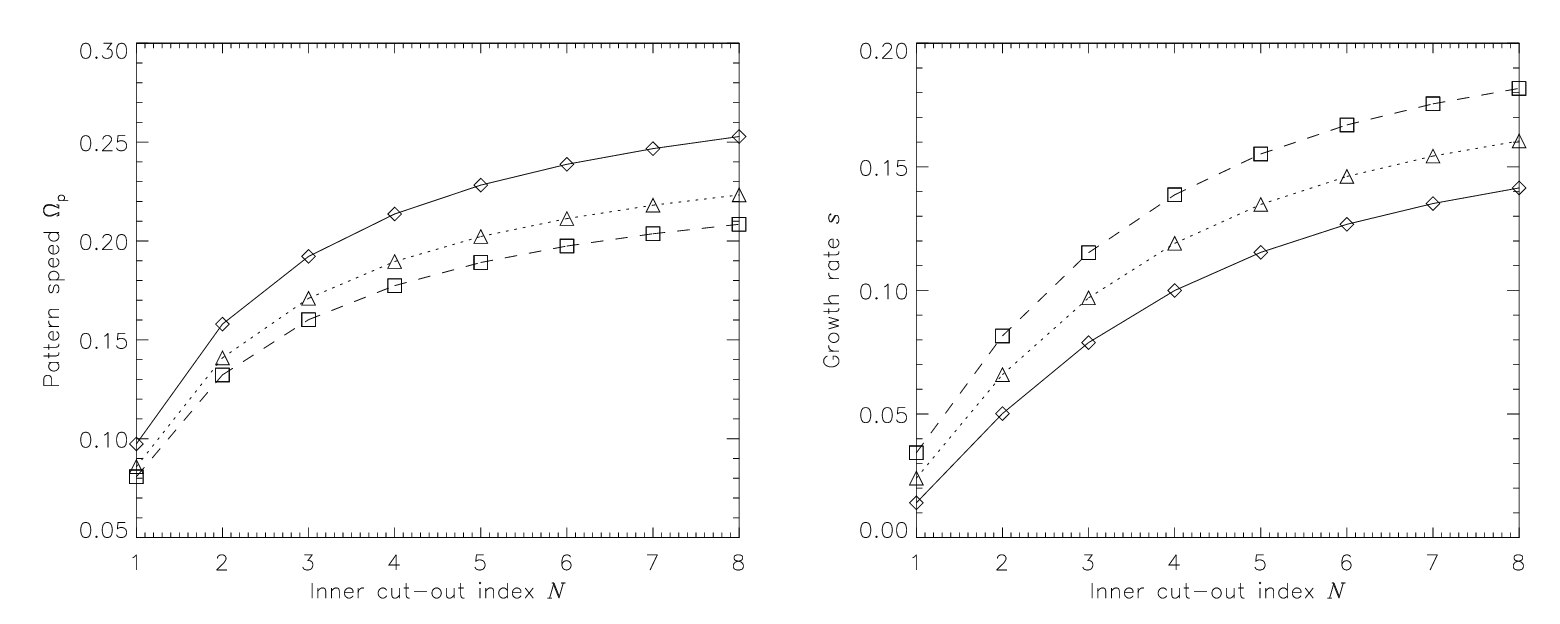,width=\textwidth,height=0.4\textwidth}
		\caption [The fastest-growing modes with $m=1$,
		$\Qsing=1.0$] {The variation of the growth rate and
		pattern speed of the fastest-growing $m=1$ modes
		plotted against cut-out index $N$.  All disks have
		$\Qsing=1.0$. The solid line marked with diamonds is
		$\beta=0.25$; the dotted line with triangles is
		$\beta=0.00$; the dashed line with squares is
		$\beta=-0.25$.}  \end{center}
\end{figure}

\begin{figure}
	\begin{center}
		\epsfig{file=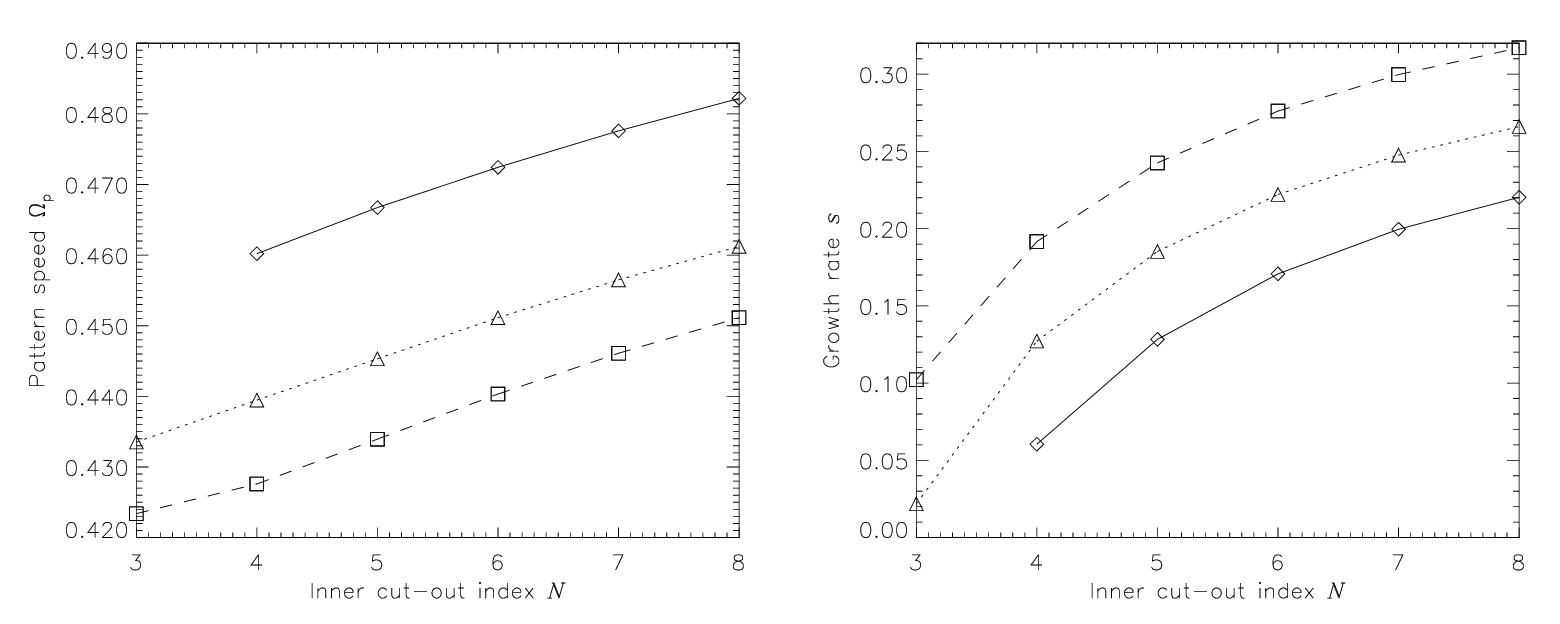,width=\textwidth,height=0.4\textwidth}
		\caption [The fastest-growing modes with $m=2$,
		$\Qsing=1.0$] {The variation of the growth rate and
		the pattern speed of the fastest-growing $m=2$ modes
		plotted against cut-out index $N$. All disks have
		$\Qsing=1.0$. The solid line marked with diamonds is
		$\beta=0.25$; the dotted line with triangles is
		$\beta=0.00$; the dashed line with squares is
		$\beta=-0.25$.}  \end{center}
\end{figure}